\documentclass[10pt]{extarticle}
\usepackage{fullpage}
\usepackage[english]{babel}
\usepackage[utf8]{inputenc}
\usepackage{amsmath, amssymb, bm}
\usepackage{graphicx}
\usepackage{subcaption}
\usepackage[colorinlistoftodos]{todonotes}
\usepackage{indentfirst}

\newcommand*{\R}{{\mathbb R}}

\newcommand*{\rd}{{\mathrm d}}

\newcommand{\hiddenpower}[2] { \ifnum \numexpr#2=1 #1 \else #1^#2 \fi }
\numberwithin{equation}{section}

\usepackage{xparse}
\ExplSyntaxOn
\newcounter{diff_order}
\newcounter{diff_power}

\newcommand{\rawdiff}[3]
{
	\setcounter{diff_order}{0}
	\clist_map_inline:nn{#3}{\stepcounter{diff_order}}
	
	\frac{\hiddenpower{#1}{\thediff_order} #2}
	{
		\def\old_var{DefaultValue}
		\setcounter{diff_power}{0}
		
		\clist_map_inline:nn{#3}
		{
			\def\new_var{##1}
			\ifnum \thediff_power=0
				\stepcounter{diff_power}
			\else
				\tl_if_eq:NNTF \new_var \old_var
				{\stepcounter{diff_power}}
				{
					#1 \hiddenpower{\old_var}{\thediff_power}
					\setcounter{diff_power}{1}
				}
			\fi

			\def\old_var{##1}
		}
		
		#1 \hiddenpower{\old_var}{\thediff_power}
	}
}

\ExplSyntaxOff

\newcommand{\lb}{\left(}
\newcommand{\rb}{\right)}

\renewcommand{\sin}[2][1]{\hiddenpower{\text{sin}}{#1} \lb #2 \rb}

\renewcommand{\cos}[2][1]{\hiddenpower{\text{cos}}{#1} \lb #2 \rb}

\renewcommand{\sinh}[2][1]{\hiddenpower{\text{sinh}}{#1} \lb #2 \rb}
\renewcommand{\cosh}[2][1]{\hiddenpower{\text{cosh}}{#1} \lb #2 \rb}

\renewcommand{\coth}[2][1]{\hiddenpower{\text{coth}}{#1} \lb #2 \rb}

\begin{document}


\begin{center}
\strut\hfill



\noindent {\LARGE{\bf{Stochastic analysis $\&$ discrete quantum systems}}}\\
\vskip 0.3in

\noindent {\footnotesize {{ANASTASIA DOIKOU, SIMON J.A. MALHAM AND ANKE WIESE}}}
\vskip 0.4in

\noindent {\footnotesize School of Mathematical and Computer Sciences, Heriot-Watt University,\\
Edinburgh EH14 4AS, United Kingdom}


\vskip 0.1in

\noindent {\footnotesize {\tt E-mail: a.doikou@hw.ac.uk, s.j.a.malham@hw.ac.uk, a.wiese@hw.ac.uk }}\\

\vskip 0.60in

\end{center}

\begin{abstract}
\noindent We explore the connections between the theories of stochastic analysis  and discrete quantum mechanical systems. Naturally these connections include
the Feynman-Kac formula, and the Cameron-Martin-Girsanov theorem. More precisely, the notion of the quantum canonical transformation is employed for computing the time propagator,  
in the case of generic dynamical  diffusion coefficients. Explicit computation of the path integral leads to a universal expression for the associated measure regardless of the form of 
the diffusion coefficient and the drift. This computation also reveals that the drift plays the role of a super potential in the usual super-symmetric quantum mechanics sense. 
Some simple illustrative examples such as the Ornstein-Uhlenbeck process and the multidimensional Black-Scholes model are also discussed. Basic examples of quantum integrable 
systems such as the quantum discrete non-linear hierarchy (DNLS) and the XXZ spin chain are presented providing specific connections between quantum (integrable) 
systems and stochastic differential equations (SDEs). 
The continuum limits of the SDEs for the first two members of the NLS hierarchy turn out to be the stochastic transport and the stochastic heat equations respectively.
The quantum Darboux matrix for the discrete NLS is also computed as a defect matrix and the relevant SDEs are derived. 
\end{abstract}

\date{}
\vskip 0.4in



\section{Introduction}

\noindent Let $\hat L_0$ be a generic second order differential operator, and  suppose $f({\mathrm x},t) $ satisfies the time evolution equation:
\begin{eqnarray}
&&-\partial_t f({\mathrm x},t) = \hat L f({\mathrm x},t)= \Big (\hat L_0 + u({\mathrm x})\Big ) f({\mathrm x},t), \label{gen0}\\
&& \hat L_0 = {1 \over 2}\sum_{i, j=1}^M g_{ij}({\mathrm x}) {\partial^2\over \partial{{\mathrm x}_i} \partial{{\mathrm x}_j}} + 
\sum_{j=1}^M b_j({\mathrm x}){\partial \over \partial{{\mathrm x}_j} } , 
~~~~~~g({\mathrm x}) = \sigma({\mathrm x})\sigma^T({\mathrm x})
 \label{general}
\end{eqnarray}
where the diffusion matrix $g({\mathrm x})$ and the matrix $\sigma({\mathrm x})$ are in general dynamical (depending on the fields ${\mathrm x}_j$) $M \times M$ matrices, 
while ${\mathrm x}$ and the drift $b({\mathrm x})$ are $M$ vector fields with components ${\mathrm x}_j$, $b_j$ respectively, and $^T$ denotes usual transposition. 
Within the quantum mechanics framework the $\hat L$ operator may be seen as the Hamiltonian of a system of $M$-interacting particles and the time evolution problem (\ref{general}) 
can be thought of as a generalized imaginary time Schr\"{o}dinger's equation. 
Feynman's path integral is a solution of (\ref{general}) that can be explicitly calculated via a time discretization scheme, and can be physically interpreted as the probability 
of the system to progress from an initial state configuration  ${\mathrm x}_0$ at time $t_0$ to a final configuration ${\mathrm x}_f$ at time $t_f$. In the statistical physics language 
the imaginary time propagator provides the partition function of a given statistical system at temperature $T\sim {1\over t}$ (see for instance \cite{Stat-Book}). 

In the quantum mechanical  set up the time evolution problem (\ref{general}) is the commencing point, while from the stochastic analysis perspective the key object is 
a given stochastic differential equation (SDE) of the form
\begin{equation}
d{\mathrm x}_t = b({\mathrm x}_t) dt + \sigma({\mathrm x}_t) d{\mathrm w}_t, \label{st1}
\end{equation}
which yields the generator $\hat L_0$ of the stochastic process and the corresponding time evolution problem via It\^{o}'s formula. 
The SDE (\ref{st1}) {\it a priori}  determines the probabilistic measure 
in the Feynman-Kac formula, which in turn provides the solution of the time evolution problem. ${\mathrm w}$ is also an $M$-vector 
with components ${\mathrm w}_j$, which are $M$-independent Wiener 
processes.  Let us now recall the definition of the Wiener process (Brownian motion)
(see e.g.  \cite{Baudoin}-\cite{PitmanYor}):

\begin{itemize}
\item {\tt Wiener process}

The one dimensional (or scalar) Wiener process ${\mathrm w}_t$ is a stochastic process with the following properties:
\begin{enumerate}
\item For $s<t$ the increment ${\mathrm w}_t - {\mathrm w}_s$ is a Gaussian with mean zero ${\mathbb E}\big ({\mathrm w}_t -{\mathrm w}_s\big ) =0$,
and variance ${\mathbb E}\big (({\mathrm w}_t -{\mathrm w}_s)^2\big ) = t-s$. Moreover, 
the increments associated to disjoint intervals are independent.

\item  ${\mathrm w}_t$ is a continuous function 

\item The process starts at $t=0$, i.e. ${\mathrm w}_0 =0$.
\end{enumerate}
In the general case of $M$-dimensional Wiener processes ${\mathrm w}_t$ is a vector field ${\mathrm w}_t = \big ({\mathrm w}_{t1}, \ldots  {\mathrm w}_{tj} \big)$; the increments 
${\mathrm w}_{tj }- {\mathrm w}_{sj}$ are Gaussians with mean zero and
${\mathbb E}\big (({\mathrm w}_{ti} -{\mathrm w}_{si})({\mathrm w}_{tj} -{\mathrm w}_{sj})\big ) = (t-s) \delta_{ij}$, which loosely speaking (in the sense of averaging or as a scaling argument) 
can be read as: $d{\mathrm w}_{ti} d{\mathrm w}_{tj} = \delta_{ij}dt$.

Note that  although ${\mathrm w}_t$ is continuous, is  nowhere differentiable. 
\end{itemize}
As pointed out the $\hat L_0$ operator and the associated time evolution equation can be obtained via It\^{o}'s formula, which is briefly reviewed below:
\begin{itemize}
\item {\tt It\^{o}'s formula}

Consider a function $f({\mathrm x}, t)$, 
where ${\mathrm x}$ satisfies (\ref{st1}) then
\begin{equation}
df({\mathrm x}, t) ={\partial f \over \partial t} dt + \sum_j {\partial f \over \partial {\mathrm x}_j}d{\mathrm x}_j + {1 \over 2 }\sum_{i,j} {\partial^2 f \over \partial {\mathrm x}_i 
\partial {\mathrm x}_j}d{\mathrm x}_id{\mathrm x}_j. \label{var}
\end{equation}
Notice the appearance of the last term when formulating the differential above. Indeed,  due to the fact that $d{\mathrm x}_{tj}$ satisfy the SDE (\ref{st1}), 
and $d{\mathrm w}_{ti} d{\mathrm w}_{tj} = \delta_{ij} dt $, the last term  is not negligible. Using (\ref{st1}) and the scaling argument for the Wiener processes, expression (\ref{var})  becomes
\begin{equation}
df({\mathrm x}, t) =\Big (\partial_t+\hat L_0 \Big )f({\mathrm x}, t)dt + \sum_{i,j} \sigma_{ij}({\mathrm x}){\partial f({\mathrm x}, t) \over\partial  {\mathrm x}_i}d{\mathrm w}_j  \label{scale1}.
\end{equation}
providing the generator $\hat L_0$ (\ref{general}), which can be thought of as a quantum mechanical Hamiltonian 
describing $M$ interacting particles, or alternatively as an $M$-dimensional quantum mechanical Hamiltonian. 
The equation $\partial_t + \hat L_0=0 $ is obtained after taking the expectation value of (\ref{scale1}).
For a rigorous proof of  (\ref{scale1}) and detailed discussions on stochastic differential equations and in particular on It\^{o}'s lemma and It\^{o}'s calculus, Feynman-Kac formula, and
Wiener processes we refer the interested reader to \cite{Baudoin}-\cite{PitmanYor}. 

\end{itemize}

After the brief review on the time evolution problem from the stochastic point of view we may now outline our perspective, and present  the main objectives of the present analysis. 
The first main objective here is to solve the generic time dependent Schr\"odinger equation (time evolution equation) (\ref{gen0}), by employing the path integral formulation.
Then from the solution of the time evolution problem via the path integral (time discretization) formulation we will arrive at the SDE  (\ref{st1}), 
and the Feynman-Kac formula (see also \cite{Simon} and references therein).  It is clear that the generic operator $\hat L_0$ is not in general self-adjoint (Hermitian), therefore 
we should also introduce the adjoint operator defined for any suitable function $f({\mathrm x}, t)$ as
\begin{eqnarray}
\hat L_0^{\dag}f({\mathrm x}, t) ={1 \over 2}\sum_{i, j=1}^M  {\partial^2\over \partial{{\mathrm x}_i} \partial{{\mathrm x}_j}}\Big (g_{ij}({\mathrm x}) f({\mathrm x}, t)\Big ) - 
\sum_{j=1}^M {\partial \over \partial{{\mathrm x}_j}}\Big (b_j({\mathrm x}) f({\mathrm x}, t)\Big ).\label{adjoint}
\end{eqnarray}
Hence, two distinct time evolution equations emerge:
\begin{itemize}
\item {\tt Fokker-Plank equation} 

For $ t_1\geq t_2$ the Fokker-Plank or Kolmogorov forward equation is given by
\begin{equation}
\partial_{t_1} f({\mathrm x},t_1) = \hat L_0^{\dag} f({\mathrm x},t_1) \label{fokker} 
\end{equation}
with known initial condition $f({\mathrm x}, t_2)=f_{0}({\mathrm x})$. 

\item {\tt Kolmogorov backward equation} 

For $t_2 \leq t_1$ the Kolmogorov backward equation is given by
\begin{equation}
-\partial_{t_2} f({\mathrm x},t_2) = \hat L_0 f({\mathrm x},t_2) \label{back} \nonumber
\end{equation}
with known final condition $f(x, t_1)=f_{f}(x)$.
\end{itemize}

To solve the PDEs above we will follow as already mentioned the path integral formulation.
We should note however, that formulating the path integral in the case of a generic $\hat L$  operator with non-constant (dynamical) diffusion coefficients  represents a particular difficulty.
More precisely, one has to deal with the situation of a non-Gaussian measure when computing the path integral in the standard way. To circumvent this complication in section 2 we employ the notion of 
\emph{quantum canonical transformation} and reduce the operator $\hat L$ to a much simpler form with constant diffusion coefficients, but with an induced/effective drift. 
Moreover, by considering both $\hat L$ and a ``complementary'' operator $\hat L^s$ --it will be defined later in the text-- we show that the drift  plays the role of a 
super-potential, whereas $\hat L$,  $\hat L^s$  may be thought of as ``super-partners''  in the typical super-symmetric quantum mechanics sense (see also e.g. 
\cite{Getzler, SUSY} and references therein). Indeed, (\ref{general}) may be seen as a generalized imaginary time Schr\"{o}dinger equation,
\begin{equation}
\hat L= \sum_{j=1}^M {\mathrm D}_{{\mathrm y}_j}^2 + V({\mathrm y}) \nonumber
\end{equation}
where ${\mathrm D}_{{\mathrm y}_j}$ is a covariant derivative and $V({\mathrm y})$ is some induced (effective) potential. At the level of the associated SDEs the existence of two 
equivalent SDEs corresponding to $\hat L$ in (\ref{general}) and its simplified version with constant diffusion coefficient respectively is shown in subsection 2.2. 
Then Feynman's path integral is evaluated for the reduced $\hat L\ $, $\hat L^s$ operators in subsection 3.1. and the findings obtained via the quantum canonical transform are confirmed.
Moreover, direct connections with Girsanov's theorem \cite{Girsanov} as well as super-symmetric quantum mechanics and gauge theories are made. The use of a precise time discretization scheme together with scaling arguments are essential for these connections.  

We  also compute explicitly the generic path integral with dynamical diffusion coefficients in subsection 3.2. by requiring that the fields involved satisfy discrete time SDEs.
This key assumption, which is natural in the context of the time discrete scheme for computing the propagator, leads to a straightforward computation of the corresponding probabilistic measure, 
that turns out to be an infinite product of Gaussians. Hence, the canonical transform implemented 
at the level of PDEs is also confirmed at the microscopic level of the path integral. It is worth emphasizing that  in this framework solutions of the underlying SDEs are required in order 
to compute time expectation values (see also \cite{Cardy}).  This formulation can be seen as an alternative to the customary  Lagrangian formulation and perturbative methods 
used in quantum and statistical theories. Various relevant examples are considered in section 4, where the use of the standard semi classical approach and the stochastic approach are displayed.
Typical examples such as  the Ornstein-Uhlenbeck process, and the multidimensional Black-Scholes model are presented, and the use of the Feynman- Kac formula in the problem of  quantum quenches is
also discussed. The computation of  the harmonic oscillator propagator (Mehler's formula) by means of stochastic analysis arguments is also presented.

Our second main objective is to explore links between SDEs, quantum integrable systems and the Darboux-B\"acklund transformation \cite{Sklyanin, Korff, DoikouFindlay}, as discussed in section 5. 
To illustrate these associations between discrete quantum systems and SDEs we discuss typical exactly solvable discrete quantum systems, such as the Discrete 
non-linear Schr\"odinger hierarchy, in particular the first two non- trivial members i.e. the Discrete-self-trapping (DST) model and the discrete NLS (DNLS) model  \cite{Eilbeck, KunduRagnisco},
as well as  the Heisenberg (XXZ) quantum spin chain \cite{Bethe, Faddeev, Korepin}.  Interesting connections between the DST model and the Toda chain are shown, 
and the SDEs associated to the DST model are solved using integrator factor techniques. We also consider both DST and DNLS models at the continuum limit, 
and we explicitly show that the associated SDEs turn to well known solvable SPDEs i.e. the stochastic transport and the stochastic heat  equation respectively. 
The stochastic heat equation in particular can be further mapped to the stochastic viscous Burgers equation (see for instance \cite{Corwin} and references therein).  
Moreover, the DST model in the presence of local defects is studied, 
the quantum Darboux-B\"acklund relations are derived for the defect and the corresponding SDEs are also identified. 
The algebraic picture is also engaged leading to a modified set of SDEs.

\section{The quantum canonical transformation}

\noindent 
As has been pointed out the formulation of  the path integral solution in the case of a generic $\hat L$   
operator with non-constant (dynamical) diffusion coefficients  represents a particular difficulty, 
given that its explicit computation leads to a non-Gaussian measure.
The main objective in this section is the transformation of the dynamical diffusion matrix in (\ref{general}) 
into identity at the level of the PDEs. Then in the next subsection motivated by the result at the PDE level we shall be able to show the corresponding 
equivalence at the SDE level generalizing what is known as the Lamperti transform (see e.g.  \cite{Karatzas, Oksendal, MM, Lamperti}). 
This result will be then utilized in section 3 for the explicit computation of the general
path integral, and the derivation of the Feynman-Kac formula.
 
Indeed, we will show in what follows that the general $\hat L$ operator can be brought into the less involved form:
\begin{equation}
\hat L = {1\over 2} \sum_{j=1}^M {\partial^2 \over \partial{{\mathrm y}_j}^2} +
\sum_{j=1}^M  \tilde b_j({\mathrm y}) {\partial \over \partial{{\mathrm y}_j}}+u({\mathrm y}) \label{constant}
\end{equation}
with an induced drift $\tilde b({\mathrm y})$. This can be achieved via a simple change of the parameters ${\mathrm x}_j$, which in the Riemannian geometry sense is nothing but a change of frame. 
Indeed, let us introduce a new set of parameters ${\mathrm y}_j$ such that:
\begin{equation}
 d{\mathrm y}_i = \sum_j \sigma^{-1}_{ij}({\mathrm x})\ d{\mathrm x}_j, ~~~~\det \sigma \neq 0, \label{frame}
\end{equation}
then $\hat L$ can be expressed in the form (\ref{constant}), and the induced drift components are given as
\begin{equation}
\tilde b_k({\mathrm y}) = \sum_j \sigma_{kj}^{-1}({\mathrm y}) b_j({\mathrm y}) +{ 1\over 2} \sum_{j,l} \sigma_{jl}({\mathrm y}) \partial_{{\mathrm y}_l} \sigma_{kj}^{-1}({\mathrm y}). \label{drifta}
\end{equation}
Bearing also in mind that $\sum_j \sigma_{jl} \sigma_{kj}^{-1} = \delta_{kl}$,
we can write in the compact vector/matrix notation:
\begin{equation}
\tilde b({\mathrm y}) = \sigma^{-1}({\mathrm y})\Big(b({\mathrm y}) -{ 1\over 2}  (\nabla_{{\mathrm y}} \sigma^T({\mathrm y}))^T\Big ), ~~~~~~
\nabla_{\mathrm y} = \Big (\partial_{{\mathrm y}_1}, \ldots, \partial_{{\mathrm y}_M} \Big ) \label{drift1}
\end{equation}
where one first solves for ${\mathrm x} ={\mathrm x}({\mathrm y})$ via (\ref{frame}).

The latter statement can be seen in a more general algebraic (Hamiltonian) frame, which is more relevant for our purposes here, especially when considering the $M$-dimensional case, 
$M \to \infty$ i.e. in a 1+1 field theoretical setting. Let $X_j,\ P_j$ be elements of the Heisenberg-Weyl algebra:
\begin{eqnarray}
&&\big [ X_i,\ P_j \big ] ={\mathrm i} \delta_{ij}. \label{comm}  
\end{eqnarray}
where ${\mathrm i} = \sqrt{-1}$.
The contribution of the commutator above gives rise to the second term in the induced drift (\ref{drift1}), and this is clearly a purely ``quantum'' effect. We shall be able in fact to reproduce 
this quantum effect via the computation of Feynman's path integral, and also we will be able to extract it at a fundamental SDE level. In the generic algebraic framework $\hat L$ is then expressed as:
\begin{equation}
\hat L( {\mathrm X},\ {\mathrm P}) =- {1 \over 2}\sum_{i, j}g_{ij}( {\mathrm X})  P_i P_j +  {\mathrm i} \sum_jb_j( {\mathrm X}) P_j +u({\mathrm X}). \label{l1}
\end{equation}
The elements ${\mathrm X},\ {\mathrm P}$ are $M$ column vectors with components $ X_j,\ P_j$, which in our setting are represented as:
\begin{equation}
 X_j \mapsto {\mathrm x}_j, ~~~~~~ P_j \mapsto - {\mathrm i}\partial_{{\mathrm x}_j}. \nonumber
\end{equation}

Now consider the following quantum canonical transformation $\hat L( {\mathrm X},\  {\mathrm P}) \mapsto \hat L( {\mathrm Y},\ {\mathrm {\cal P }})$:
\begin{eqnarray}
Y_i  = \sum_j \int^{X_j} \sigma^{-1}_{ij}({\mathrm X}')\ d X'_j, ~~~~~~ {\cal P}_i = \sum_j \sigma_{ij}({\mathrm X})\ P_j \label{canon} 
\end{eqnarray}
where in the integral above and the derivative in (\ref{comm}) $X_j$ are formally treated as  parameters. After the canonical transformation (\ref{canon}) is implemented the operator $\hat L$ reads as
\begin{equation}
\hat L( {\mathrm Y},\ {\mathrm  {\cal P}}) =- {1 \over 2}\sum_j  {\cal P}_j^2 + {\mathrm i} \sum_j \tilde b_j({\mathrm Y})\   {\cal P}_j+u({\mathrm Y}) \label{l2}
\end{equation}
and naturally $Y_j,\ {\cal P}_j$ are also elements of the Heisenberg-Weyl algebra (\ref{comm}) represented as:
\begin{equation}
Y_j \mapsto {\mathrm y}_j, ~~~~~ {\cal P}_j \mapsto -{\mathrm i}\partial_{{\mathrm y}_j} \label{diff2}
\end{equation}
This procedure is analogous to the quantum Darboux-B\"{a}cklund transformation studied in \cite{DoikouFindlay}; similar results are presented in \cite{Sklyanin, Korff}, 
where the so-called Baxter's $Q$-operator \cite{Baxter} is used as the generating function of the quantum canonical transformation.

We have established via the quantum canonical transform that the general diffusion matrix can turn to the identity. Let us now consider both $L$ and
\begin{equation}
\hat L^s= -{1\over 2} {\mathrm {\cal P}}^T {\mathrm {\cal P} }+i {\mathrm  {\cal P}}^T \tilde b({\mathrm Y}) + u({\mathrm Y}). \label{super1}
\end{equation}
$\hat L^s$ not to be confused with the adjoint operator $\hat L^{\dag}$. The two operators coincide only when a purely imaginary drift $\tilde b$ is considered. 

\begin{itemize}
\item {\tt Remark 1: the drift as super-potential}

We introduce the following compact notation: $\forall F, \ \mbox{define}\ F^D \in \{F,\ F^s\}$. We can bring the $\hat L^D$ operators to an even more familiar form 
by setting: ${\mathrm  {\mathbb P}}= {\mathrm {\cal P}} - i\tilde b({\mathrm Y})$, where $Y_j,\ {\mathbb P}_j$ are also elements of the Heisenberg-Weyl algebra (\ref{comm}). Then
\begin{eqnarray}
&& \hat L^D = {1\over 2} {\mathrm  {\mathbb P}}^T {\mathrm  {\mathbb P}}  + V^D({\mathrm Y}) \label{0} \nonumber\\
&& V^D({\mathrm Y})  = - {1 \over 2} \tilde b^T({\mathrm Y}) \tilde b({\mathrm Y}) - {1 \over 2} \nabla_{\mathrm Y} \tilde b({\mathrm Y}) + u^D({\mathrm Y}) \label{1} \\
&& u^s({\mathrm Y})= u({\mathrm Y})+ \nabla_{\mathrm Y} \tilde b({\mathrm Y}). \label{2} \nonumber
\end{eqnarray}
where $V^D$ are effective potentials, produced exclusively by the drift. In terms of differential operators then one obtains via:
\begin{eqnarray}
\hat L^D =  {1\over 2} \sum_j  {\mathrm D}_{{\mathrm y}_j}^2 + V^D({\mathrm y}) \label{covariant} 
\end{eqnarray}
where $ {\mathrm D}_{{\mathrm y}_j}=\partial_{{\mathrm y}_j} + \tilde b_j({\mathrm y})$ is the covariant derivative. Interestingly $\tilde b$ turns out to be the super-potential that produces 
$V, \ V^s$ and satisfies the familiar Riccati equations (\ref{1}).  The statements at the algebraic level will be explicitly confirmed via the evaluation of the path integrals associated 
to both operators $\hat L,\ \hat L^s$.

\item {\tt Remark 2: Riccati equation $\&$ Girsanov's theorem}

Note that the Riccati equation (\ref{1}) emerging from the identification of the effective potential due to the presence of the drift, -and hence the covariant derivative 
(\ref{covariant})- is essentially 
the origin of Girsanov's theorem on the change of probabilistic measure (we refer the interested reader to \cite{Baudoin}-\cite{PitmanYor}) from the PDE point of view. 
This will be even more transparent in the next section from the explicit path integral computation.

\end{itemize}

\subsection{Equivalence of SDEs}

\noindent 
The generator $\hat L_0$ (\ref{general}) as already discussed can be obtained via It\^{o}'s lemma from the SDE (\ref{st1}), 
while the operator in the simpler equivalent form established previously:
\begin{eqnarray}
{\hat L}_0 = {1\over 2} \sum_{j=1}^M {\partial^2 \over \partial{{\mathrm y}_j}^2} +
\sum_{j=1}^M  \tilde b_j({\mathrm y}) {\partial \over \partial{{\mathrm y}_j}} \label{simple}
\end{eqnarray}
is obviously obtained via the following simpler SDE
\begin{equation}
d {\mathrm y}_t = \tilde b({\mathrm y}_t) dt + d{\mathrm w}_t, \label{st2}
\end{equation}
where recall $\tilde b$ is defined in (\ref{drifta}), (\ref{drift1}).

The main aim now is to show that these two SDEs (\ref{st1}) and (\ref{st2}) are in fact equivalent. Indeed, let us start from (\ref{st2}), multiply both sides of the equation 
with $\sigma({\mathrm y}_t)$ and add It\^{o}'s correction  i.e.
\begin{equation}
\sigma({\mathrm y}_t)  d {\mathrm y}_t   +{1\over 2} d\sigma({\mathrm y}_t)  d{\mathrm y}_t= \sigma({\mathrm y}_t) \tilde b({\mathrm y}_t) dt  +
{1\over 2} d\sigma({\mathrm y}_t) d{\mathrm y}_t+ \sigma({\mathrm y}_t) d{\mathrm w}_t. \label{inter}
\end{equation}
Recall the connection between It\^{o} and Stratonovich calculus:
\begin{eqnarray}
f({\mathrm y}_t)\circ d {\mathrm y}_t = f({\mathrm y}_t)d{\mathrm y}_t +{1\over 2}df({\mathrm y}_t)d{\mathrm y}_t ~~~~
\mbox{and} ~~~d{\mathrm y}_i d{\mathrm y}_j= \delta_{ij} dt \label{correction}
\nonumber
\end{eqnarray}
where $f({\mathrm y}_t)\circ d {\mathrm y}_t$ refers to Stratonovich  calculus, and $ f({\mathrm y}_t) d {\mathrm y}_t$ to It\^{o}. 
More comments on this relation will be given in the next subsection, where the discrete time scheme is discussed via the 
path integral computation. Recall also the definition of the modified drift $\tilde b$ (\ref{drifta}):
\begin{equation}
\sigma({\mathrm y}_t) \tilde b({\mathrm y}_t) dt  +{1\over 2} d\sigma({\mathrm y}_t) d{\mathrm y}_t = b({\mathrm y}_t)dt, \label{corr2} \nonumber
\end{equation}
then (\ref{inter}) becomes
\begin{equation}
\sigma({\mathrm y_t}) \circ d {\mathrm y}_t = b({\mathrm y}_t) dt + \sigma({\mathrm y}_t)d{\mathrm w}_t. \nonumber
\end{equation}
We now set
\begin{equation}
d{\mathrm x}_t = \sigma({\mathrm y}_t) \circ d {\mathrm y}_t \label{trans} 
\end{equation}
and we immediately recover the original SDE
\begin{equation}
d{\mathrm x }_t= b({\mathrm x}_t) dt + \sigma({\mathrm x}_t) d{\mathrm w}_t \nonumber
\end{equation}
after solving ${\mathrm y} = {\mathrm y}({\mathrm x})$ through (\ref{trans}). Integration of (\ref{trans}) in the Stratonovich calculus frame -- i.e. the usual calculus rules apply-- yields
\begin{equation}
{\mathrm x}_i  = \sum_j \int^{{\mathrm y}_j} \sigma_{ij}({\mathrm y}')\circ d {\mathrm y}'_j, \nonumber
\end{equation}
and the latter as expected is nothing but the inverse of (\ref{canon}).

Naturally we could have started from (\ref{st1}) and ended up to (\ref{st2}) via the reverse process. Indeed, we start from (\ref{st1}) we multiply the equation by $\sigma^{-1}$, add It\^{o}'s correction  
${1\over 2}d\sigma^{-1}({\mathrm x})d{\mathrm x}$ at both sides of the equation, and use  (\ref{correction}). Then setting $d{\mathrm y} =\sigma^{-1}({\mathrm x})\circ d{\mathrm x}$ we end up to 
the desired equation (\ref{st2}). With this we conclude of our proof on the equivalence of the two SDEs (\ref{st1}) and (\ref{st2}). Relevant results on the change of the diffusion matrix 
from the SDE point of view via the so called Lamperti transform are presented in \cite{MM}, (regarding the Lamperti transform we refer the interested reader to \cite{Karatzas, Oksendal, Lamperti}).

\section{The path integral}

\subsection{Identity diffusion matrix}

\noindent  We have established in the previous section that if the generic dynamical diffusion matrix is invertible then $\hat L$ can be brought into a simple 
form with identity diffusion matrix via the quantum canonical transformation. We basically confirm the results of  section 2 via the explicit computation of the relevant path integral 
using the Trotter product formula (see for instance \cite{Simon}).  We review in this subsection the computation of the path integral in the case when the 
diffusion matrix is constant, but in the presence of a general drift. We explicitly compute in what follows the propagator associated to the operator $\hat L$ 
using the standard discrete time frame and make various interesting connections.  More precisely: 
\begin{enumerate}
\item We make a direct connection with the discrete time version of 
Cameron-Martin-Girsanov theorem on the change of the probabilistic measure, which is also associated to the presence of a geometric phase in gauge theories.
\item We show that the path integral can be alternatively computed 
by assuming the existence of underlying discrete time SDEs. This fundamental assumption leads to a straightforward computation of the associated measure, 
which turns out to be a Gaussian. This result is generalized in the next subsection where dynamical diffusion coefficients are also considered. 
\item We show that the drift, which is essentially a gauge field plays the role of the super-potential with $\hat L$, $\hat L^s$ being 
super-partners in the usual super-symmetric 
quantum mechanics sense, confirming the findings of the previous section. 
\end{enumerate}

Our starting point is the time evolution equation (\ref{adjoint}), (\ref{fokker}),  (\ref{constant}):
\begin{equation}
\partial_t f({\mathrm y} ,t) = \hat L^{\dag}  f({\mathrm y}, t), \label{evolution} \nonumber
\end{equation}
we then explicitly compute the propagator $K({\mathrm y}_f, {\mathrm y}_i| t, t')$:
\begin{eqnarray}
f({\mathrm y} ,t) &=& \int  \prod_{j=1}^M d y_j'\ K({\mathrm y}, {\mathrm y}'| t, t')f({\mathrm y}' ,t') \label{kernel0}\\
&=& \int \prod_{n=1}^N \prod_{j=1}^M d y_{jn}\ 
\prod_{n=1}^N K({\mathrm y}_{n+1}, {\mathrm y}_n| t_{n+1}, t_n)f({\mathrm y}_1 ,t_1). \label{kernel}
\end{eqnarray}
In what follows we shall reproduce It\^{o}'s correction as the result of quantum/statistical fluctuations --i.e. use of suitable scaling arguments-- 
and also derive the corresponding SDEs via the exact computation of the propagator $K$, assuming 
no a priori knowledge of stochastic analysis. Within this frame the discrete time analogue of Girsanov's theorem is a straightforward  result.

Employing the standard time discretization scheme or the semi-group property, as shown above, (see also for instance \cite{Simon}) 
the path integral or time propagator can be expressed as
\begin{eqnarray}
&& K({\mathrm y}_f, {\mathrm y}_i| t, t') =  \nonumber\\
&& {1 \over (2\pi)^{NM}} \int d{\bf y}\  d {\bf  p}\  \exp \Big [ -{\delta \over 2} \sum_{n=1}^N \sum_{j=1}^M  p^2_{jn} + 
{\mathrm i} \sum_{n=1}^N \sum_{j=1}^M p_{jn}\big (\Delta {\mathrm y}_{jn} - 
\delta\ \tilde  b_{jn}({\mathrm y})\big )+\delta\sum_{n=1}^N  u_n({\mathrm y})\Big ]  \nonumber\\
&&\label{before}
\end{eqnarray}
where we define
\begin{eqnarray}
d{\bf y} = \prod_{n=2}^N\prod_{j=1}^M d{\mathrm y}_{jn}, ~~~~~d{\bf p}= \prod_{n=1}^N\prod_{j=1}^M dp_{jn},\label{before2} \nonumber
\end{eqnarray}
where $\delta= t_{n+1}- t_{n}$ and with boundary conditions: ${\mathrm y}_f = {\mathrm y}_{N+1},  ~~ {\mathrm y}_i = {\mathrm y}_{1}, ~~t_i =t'=0$
($t'$ will be dropped henceforth for brevity), $~t_f=t$; $p_{nj}$ 
are also known as response variables.  
To obtain the latter formula we have inserted the unit $N$ times, (${1\over 2 \pi} \int d{\mathrm y}_{jn}\ dp_{jn}\ e^{ {\mathrm i} p_{jn}({\mathrm y}_{jn}-a)}=1$), 
associated to each component ${\mathrm y}_j$. 
After performing the Gaussian integrals with respect to the $p_{jn}$ parameters we conclude:
\begin{eqnarray}
&& K({\mathrm y}_f, {\mathrm y}_i| t) = \int d{\bf q}\ \exp\Big [- \sum_j \sum_{n}{\big (\Delta {\mathrm y}_{jn} -\delta \tilde b_{jn}({\mathrm y})\big)^2 \over 2\delta} +
\delta\sum_n  u_n({\mathrm y})\Big ] \label{exp}\\
&&  d{\bf q}  ={1 \over (2 \pi \delta )^{NM\over 2}} \prod_{n=2}^N \prod_{j=1}^M  d{\mathrm y}_{jn} \nonumber
\end{eqnarray}
where $f_{n} = f_n({\mathrm y}_n)$ and $\Delta {\mathrm y}_{jn} = {\mathrm y}_{jn+1} - {\mathrm y}_{jn}$.
At this stage we are dealing with an $M\times N$ space-time lattice; $n$: time indices, $j$: space indices.

We now expand the square in the exponential in (\ref{exp}), and consider the continuum time limit as follows ($N\to \infty,\ \delta \to 0$):
\begin{eqnarray}
&&  \sum_{n=1}^N \tilde b^T_{n}({\mathrm y})\Delta{\mathrm y}_{n}\ \to\ \int\tilde b^T({\mathrm y}_s) d {\mathrm y}_s  \label{ito11}\\
&& {\delta\over 2} \sum_{n=1}^N \tilde b_{n}^T({\mathrm y}) \tilde b_{n}({\mathrm y})\  \to\  {1\over 2}\int_{0}^t \tilde 
b^T({\mathrm y}_s) \tilde b({\mathrm y}_s)ds \label{ito22}\\
&&\delta\sum_{n=1}^N  u_n({\mathrm y})\ \to \ \int_0^t u({\mathrm y}_s) ds, \label{integrals}
\end{eqnarray}
where $\int_{0}^t \tilde b^T({\mathrm y}_s) d{\mathrm y}_s$ corresponds to an It\^{o} type integral, 
due to the chosen discretization scheme  above. We shall discuss the It\^o-Stratonovich 
integral  correspondence later in this subsection, as these two types of integrals emerge due to two distinct 
time discretization schemes.The propagator then takes the form
\begin{eqnarray}
 K({\mathrm y}_f,  {\mathrm y}_i| t) &=& \int d{\mathbb P}\ \exp\Big [  -{1\over 2} \int_{0}^t 
 \tilde b^T({\mathrm y}_s)\tilde b({\mathrm y}_s)ds  + \int \tilde b^T({\mathrm y}_s)d{\mathrm y}_s 
+\int_0^t u({\mathrm y}_s) ds \Big ] \label{fin0} 
\end{eqnarray}
where we define
\begin{equation}
d{\mathbb P} = \lim_{\delta \to 0} \lim_{N \to \infty } {1 \over (2 \pi \delta )^{NM\over 2}} \  
\exp\Big [-\sum_{n=1}^N\sum_{j=1}^M {(\Delta {\mathrm y}_{jn})^2\over 2\delta}\Big ]
\ \prod_{n=2}^N \prod_{j=1}^M d{\mathrm y}_{jn}.  \label{finalbb}
\end{equation}

We can follow an alternative path to compute the propagator in a straightforward way by making explicit use of the corresponding SDEs. 
Recall expression (\ref{before}) and assume that:
\begin{equation}
\Delta {\mathrm y}_n - \delta \tilde b_n({\mathrm y}) =  \Delta {\mathrm w}_n \label{DSDE0}
\end{equation}
assuming also that ${\mathrm w}_n$ are independent of the vector fields ${\mathrm y}_n$.
After performing the standard Gaussian integrals in (\ref{before}) subject to (\ref{DSDE0}) we conclude that
\begin{eqnarray}
&& K({\mathrm y}_f, {\mathrm y}_i|t) =\int  d {\mathbb M}\ e^{\int_{0}^t u({\mathrm y}_s) ds}, \label{B}\\
&& d {\mathbb M} = \lim_{\delta \to 0} \lim_{ N \to \infty}\ {1 \over (2 \pi \delta )^{NM\over 2}} \ 
\exp\Big [- {1\over 2 \delta} \sum_{n=1}^N \Delta {\mathrm w}_n^T \Delta {\mathrm w}_n \Big  ]\
 \prod_{n=2}^N \prod_{j=1}^M d {\mathrm w}_{jn}. \nonumber
\end{eqnarray}
Note that the presence of the Gaussians in the measure suggests that $\Delta{\mathrm w}_{nj}$ are normal variables with zero mean 
and $\Delta {\mathrm w}_{in} \Delta {\mathrm w}_{jn} = \delta_{ij}\delta$ (in the sense of expectation value). 
Moreover, we assume that ${\mathrm w}_{0} =0$, i.e. we are considering Wiener paths. 
This will become transparent in the next subsection when evaluating explicitly the measure for the 
general case of dynamical diffusion coefficients, using the Fourier representation of the Wiener paths. 
Expression (\ref{DSDE0}) can be then  seen as the discrete time analogue of the SDE (\ref{st2}). 
Explicit evaluation of the propagator via (\ref{fin0}) naturally leads to the use of the Lagrangian formulation, which is usually implemented in 
quantum systems and quantum/statistical field theories.  Evaluation of $K({\mathrm y}, {\mathrm y}'| t)$ on the other hand via (\ref{B}) 
requires the solution of the corresponding SDEs.

\begin{itemize}

\item {\tt Remark 3: Radon-Nikodym derivative $\&$ Girsanov's theorem}

Note that the ratio (\ref{fin0}), (\ref{B})
\begin{equation}
{d{\mathbb M}\over d{\mathbb P}} = \exp\Big [ -{1\over 2} \int_{0}^t \tilde b^T({\mathrm y}_s)  
\tilde b({\mathrm y}_s) ds  + \int \tilde b^T({\mathrm y}_s) d{\mathrm y}_s\Big ] \label{ratio} \nonumber
\end{equation}
turns out to be the Radon-Nikodym derivative in analogy to Girsanov's theorem \cite{Girsanov} on the change of measure. 
Indeed, as already pointed out there is a one to one correspondence with the results of section 2 especially in relation 
to the notions of the effective potential $V$ (\ref{1}) and the covariant derivative (\ref{covariant}).

\item {\tt Remark 4: It\^o vs Stratonovich calculus}

We can now reproduce the It\^o-Stratonovich correspondence via distinct discretization schemes. 
The main reason we address this issue here is because Stratonovich type integrals follow 
the usual calculus rules, and are the ones  that are usually employed in the path integral formulation in 
quantum/statistical physics. Let us focus again on the square of the exponential (\ref{exp}) 
and in particular on the following term:
\begin{equation}
\sum_{j=1}^M \sum_{n=1}^N \tilde b_{jn}\ \Delta {\mathrm y}_{jn} = \sum_{j=1}^M  \sum_{n=1}^N \tilde b^+_{jn}\ \Delta {\mathrm y}_{jn} - 
{1\over 2}\sum_{j=1}^M  \sum_{n=1}^N\Delta \tilde  b_{jn}\ \Delta {\mathrm y}_{jn} \label{fluctuations}
\end{equation}
where $\tilde b_{jn}^+ = {1\over 2}\big (\tilde b_{jn+1} +\tilde b_{jn}\big ) $ and $\Delta  \tilde b_{jn}= \tilde b_{jn+1} - \tilde b_{jn}$. 
We also take  into consideration the scaling rule, in the sense of averaging, $\Delta {\mathrm y}_{in}\ \Delta {\mathrm y}_{jn} \sim \delta_{ij}\ \delta$, 
and consider the continuum time limit in (\ref{fluctuations}), ($\delta \to 0, \ N \to \infty$):
\begin{eqnarray}
&& \sum_{j=1}^M \sum_{n=1}^N \tilde b_{jn}\ \Delta {\mathrm y}_{jn}\  \to\ \sum_{j=1}^M\int\tilde b_j({\mathrm y}_s) d {\mathrm y}_{sj},  ~~~~~~~~\mbox{It\^o integral} \label{is1}\\
&&   \sum_{j=1}^M \sum_{n=1}^N \tilde b^+_{jn}\ \Delta {\mathrm y}_{jn}\  \to\ \sum_{j=1}^M\int\tilde b_j({\mathrm y}_s) \circ d {\mathrm y}_{sj}, ~~~~~~\mbox{Stratonovich integral} \label{is2}\\
&& {1\over 2}\sum_{j=1}^M  \sum_{n=1}^N\Delta \tilde  b_{jn}\ \Delta {\mathrm y}_{jn}\ \to\  \sum_{j=1}^M\int_0^t \partial_{{\mathrm y}_j} \tilde b_j({\mathrm y}_{s}) ds. \label{is3}
\end{eqnarray}
The above  lead to the It\^o-Stratonovich correspondence (\ref{fluctuations})-(\ref{is3}):
\begin{equation}
\int \tilde b^T({\mathrm y}_s) d{\mathrm y}_s = \int \tilde b^T({\mathrm y}_s)\circ d{\mathrm y}_s -{1\over 2}\int_0^t \nabla_{\mathrm y} \tilde b({\mathrm y}_s) ds. \label{corresp}
\end{equation}

\item {\tt Remark 5: gauge theories}

It is clear that the systems we are considering here (\ref{adjoint}), (\ref{fokker}),  (\ref{constant}) are typical gauge theories, with the vector field of the drift $\tilde b$ being the gauge field.
After taking into consideration the It\^o- Stratonovich correspondence discussed above (\ref{corresp}) we may re-express the propagator (\ref{fin0}) as:
\begin{equation}
K({\mathrm y}_f,  {\mathrm y}_i| t) = \int d{\mathbb P} \exp\Big [  -{1\over 2} \int_{0}^t  \tilde b^T({\mathrm y}_s)\tilde b({\mathrm y}_s)ds  + \int \tilde b^T({\mathrm y}_s)\circ d{\mathrm y}_s 
-{1\over 2}\int_0^t \nabla_{\mathrm y} \tilde b({\mathrm y}_s) ds+\int_0^t u({\mathrm y}_s) ds \Big ]. \label{fin02}
\end{equation}
Notice the second term in (\ref{fin0}) (i.e. the Stratonovich type integral, i.e. standard calculus rules apply)  corresponds to a standard geometric phase (see e.g. Berry phase \cite{Berry}) 
arising in gauge theories, with typical example the  Bohm-Aharonov effect \cite{Bohm}. It is clear that the first and third terms in the expression 
above can be combined with the potential $u$ to yield  an effective potential $V$, which is in agreement with the findings of section 2 and the associated 
Riccati equation (\ref{1}) for the gauge field (super-potential).

\item {\tt Remark 6:  super-symmetry $\&$ drift}

Having introduced the It\^o-Stratonovich correspondence at the microscopic level it  is worth pointing out that the propagator $K^s$ for the ``super-partner'' $L^s$ (\ref{super1}) can 
also be evaluated and one concludes that the difference ${\cal D} = K^s - K$ (\ref{fin0})  provides indeed the suitable expression in the context of super-symmetric quantum mechanics, 
related also to the index theorem (see for instance \cite{Getzler, SUSY} and references therein),
\begin{equation}
{\cal D} = 2 \int d\hat {\mathbb M}\ \exp \Big [ \int_{0}^t u({\mathrm y}_s) ds\Big ]\ \sinh{{1\over 2}\int_0^t  \nabla_{{\mathrm y}_s} \tilde b({\mathrm y}_s)ds}, \nonumber
\end{equation}
where now
\begin{equation}
{d\hat {\mathbb M}\over d{\mathbb P}} = \exp\Big [ -{1\over 2} \int_{0}^t \tilde b^T({\mathrm y}_s)  \tilde b({\mathrm y}_s) ds  + 
\int \tilde b^T({\mathrm y}_s)\circ d{\mathrm y}_s\Big ]. \label{ratio2}\nonumber 
\end{equation}
This result is also in line with the findings of section 2.

\end{itemize}

\subsection{Dynamical diffusion matrix}

\noindent
We examine the general case with dynamical diffusion matrix (\ref{gen0}), (\ref{general}), although we showed its equivalence with a case of identity diffusion 
matrix and modified drift via the quantum canonical transform.  The reason we view this case separately is because, as already mentioned, the underlying 
SDEs turn out to play a crucial role when computing the path integral. Indeed, instead of using the customary perturbative methods of computing the path 
integral one can alternatively exploit the solutions of the relevant SDEs (see also \cite{Cardy}). Specifically, our main aim in this subsection is the explicit 
computation of the propagator in the more general case of dynamical diffusion coefficients and drift. We first focus on the computation of  
the general  path integral  by employing the discrete analogue of the general SDEs (\ref{st1}) as a fundamental assumption naturally emerging after performing the relevant 
Gaussian integrals in the propagator. 

The connection with Girsanov's theorem 
and the appearance of the geometric term $\int_{0}^t \tilde b^T({\mathrm y}_s) \circ  d{\mathrm y}_s$
requires the results of section 2 and the explicit computation of the first part of the previous subsection. 
A brief outline of the general path integral computation employing the approach of the previous subsection, 
which led to the discrete time version of Girsanov's theorem, is presented in the end of the subsection

Following the prescription of the preceding subsection we can compute the corresponding path integral for the general case (\ref{general}) via (\ref{adjoint}), (\ref{fokker}) 
and express it in the compact vector/matrix notation:
\begin{equation}
K({\mathrm x}_f, {\mathrm x}_i| t) = {1 \over (2\pi)^{NM}} \int  d{\bf x}\  d {\bf  p}\ \exp \Big [-{\delta \over 2} \sum_{n=1}^N {\mathrm p}_n^{T}g_n({\mathrm x}) {\mathrm p}_n +
 {\mathrm i} \sum_{n=1}^N {\mathrm p}_n^T \big (\Delta {\mathrm x}_n - \delta b_n({\mathrm x})\big )+\delta \sum_{n=1}^N u_n({\mathrm x})\Big ] \label{discrete}
\end{equation}
where $\Delta {\mathrm x}_{n} = {\mathrm x}_{n+1} - {\mathrm x}_{n}$ and recall the diffusion matrix is $g_n = \sigma_n \sigma_n^T$.

We come now to our main assumption, let
\begin{equation}
\Delta {\mathrm x}_n - \delta b_n({\mathrm x}) = \sigma_n({\mathrm x}) \Delta {\mathrm w}_n \label{DSDE}  \nonumber
\end{equation}
and as in the previous subsection ${\mathrm w}_n$ are independent of the vector fields ${\mathrm x}_n$, moreover ${\mathrm w}_{nj}$ are independent normal variables 
and $\Delta {\mathrm w}_{in} \Delta {\mathrm w}_{jn} = \delta_{ij} \delta$. 
After performing the standard Gaussian integral:
\begin{equation}
\int d {\bf p}_n\ \exp\Big [-{\delta \over 2} {\mathrm p}^T_n g_n({\mathrm x}) {\mathrm p}_n + {\mathrm i}{\mathrm p}_n^T \sigma_n({\mathrm x}) \Delta {\mathrm w}_n\Big ] =  
 \big(\det \sigma_n({\mathrm x}) \big)^{-1}\Big ( {2\pi \over \delta}\Big )^{M\over 2} e^{- {1\over 2 \delta} \Delta {\mathrm w}_n^T \Delta {\mathrm w}_n} \label{gauss2}
\end{equation}
we conclude that
\begin{eqnarray}
&& K({\mathrm x}_f, {\mathrm x}_i|t) =\int  d {\mathbb M}\ e^{\int_{0}^t u({\mathrm x}_s) ds}, \label{discrete1} \\
&& d {\mathbb M} =  \lim_{\delta \to 0} \lim_{N \to\infty}\ {1 \over (2 \pi \delta )^{NM\over 2}}\  \exp\Big [- {1\over 2 \delta} \sum_{n=1}^N \Delta {\mathrm w}_n^T 
\Delta {\mathrm w}_n\Big ]\ \prod_{n=1}^N\big (\det \sigma_n({\mathrm x})\big )^{-1} \prod_{j=1}^M \prod_{n=2}^N d{\mathrm x}_{jn},  \label{measure11}
\end{eqnarray}
however: $(\det \sigma_n({\mathrm x}))^{-1} \prod_{j=1}^M d{\mathrm x}_{jn} =  \prod_{j=1}^M  d{\mathrm w}_{jn}$, which is a typical change of the volume element
(see also relevant change of variables discussed in section 2 equ. (2.2)). We can now explicitly evaluate the measure from the explicit expression just above 
in the continuum limit\footnote{When keeping $N$ finite we have from (\ref{discrete})
\begin{equation}
d {\mathbb M} = \big (\det \sigma({\mathrm x}_1)\big )^{-1} {1 \over (2 \pi \delta )^{M\over 2}} \  \exp\Big [- {1\over 2 \delta} \sum_{n=1}^N \Delta {\mathrm w}_1^T 
\Delta {\mathrm w}_1\Big ]\  d{\mathbb M}^-. \nonumber
\end{equation}
The factor in front of $d {\mathbb M}^-$ is nothing but the infinitesimal kernel $K({\mathrm x_2}, {\mathrm x_1} |\delta\to 0) =   (\det \sigma({\mathrm x}_1))^{-1} \delta(\Delta{\mathrm w}_1)$ 
compatible with the initial conditions of $K$. The pre-factor $  \big (\det \sigma({\mathrm x}_1)\big )^{-1} $  is essentially absorbed in the measure via the change of the volume element 
 $\big (\det \sigma({\mathrm x}_n)\big )^{-1} \prod_{j=1}^M dx_{jn} =  \prod_{j=1}^M  dw_{jn}$ .}.


To compute the corresponding measure (\ref{measure11}) (see also footnote 1)
\begin{equation}
d{\mathbb M}=\lim_{\delta \to 0} \lim_{N \to\infty}\ {1 \over (2 \pi \delta )^{NM\over 2}}\  \exp\Big [- {1\over 2 \delta} \sum_{n=1}^N \Delta {\mathrm w}_n^T 
\Delta {\mathrm w}_n\Big ]\ \prod_{j=1}^M \prod_{n=2}^N d{\mathrm w}_{jn}
\end{equation}
 in the continuum limit
we consider the
Fourier representation on $[0,\  t]$ for ${\mathrm w}_s$, i.e. Wiener's representation of the Brownian path:
\begin{equation}
{\mathrm w}_s ={ {\mathrm f}_0\over \sqrt{t} } s + \sqrt{2 \over t} \sum_{k >0} { {\mathrm f}_k\over \omega_k} \sin{ \omega_ks}, ~~~~\omega_k = {2\pi k \over t}. \label{random}
\end{equation}
${\mathrm f}_0 = {{\mathrm w}_t\over\sqrt{t}  }$ and ${\mathrm f}_k,\ k\in \{0,\ 1,  \ldots\}$ 
are $M$ vectors with components ${\mathrm f}_{kj},\ j \in \{1,\  2,  \ldots,  M\}$  being  standard normal variables.  We are interested in the computation of the measure in the continuum limit 
$N\to \infty,\ \delta \to 0$, and we also recall the following boundary conditions: ${\mathrm w}(s=0) = 0,\ {\mathrm w}(s=t) ={\mathrm w}_t$. 
Taking all the above considerations into account we conclude:
\begin{eqnarray}
&& d {\mathbb M} = {e^{-{1\over 2t}  {\mathrm w}^T_{t}{\mathrm w}_{t}} \over (2\pi t)^{{M\over 2}}}\  d{\mathbb M}_0\nonumber\\
&& d {\mathbb M}_0 =  \prod_{k \geq 1} \prod_{j=1}^M{ d{\mathrm f}_{kj} \over \sqrt {2\pi }}\  \exp[-{1\over 2} \sum_{k\geq1}\sum_j {\mathrm f}_{kj}^2]. \label{measure1}
\end{eqnarray}
The measure naturally is expressed as an infinite product of Gaussians regardless of the specific forms of the diffusion coefficients and the drift, which is indeed an elegant result. However, 
the price one pays is that solutions of the SDEs, available in general via iteration techniques, are now required. In any case this scheme can be seen as an alternative to the established
perturbative approach employed in quantum mechanical systems and quantum field theories (see also \cite{Cardy}), therefore algebraic and numerical techniques developed in solving 
SDEs will be of great relevance (see e.g. \cite{MalhamWiese, FardMalhamWiese} and references therein). The significant observation is the existence of the heat kernel in front of $d {\mathbb M}_0$. 
This boundary term plays a crucial role when computing explicitly the propagator or in general when computing time expectation values as will be transparent later in the text. 
In the more standard case where $\hat L$ is expressed as ${1\over 2} \sum_i\partial_{{\mathrm x}_i}^2 +u({\mathrm x})$ the underlying SDE is
\begin{equation}
d{\mathrm x}_t = d{\mathrm w}_t\  \Rightarrow\  {\mathrm x}_s = {\mathrm x}_0 + {\mathrm w}_s. \label{sequation} 
\end{equation}
It is clear that in more complicated cases, i.e. in the presence of drift the relation between ${\mathrm w}_t$ and ${\mathrm x}_t$ 
may become significantly involved depending on the specific form of the associated 
SDEs. A few simple illustrative examples are discussed in section 4. 

Alternatively, the problem described above can be thought of as a typical optimal control problem (feedback problem). 
More precisely, the minimization of the action $\int_{0}^t ds \big ({1 \over 2} \eta_s^T\eta_s - u({\mathrm x}_s)\big )$ should be required subject to the equation 
$\dot {\mathrm x}_t = b({\mathrm x}_t) + \sigma({\mathrm x}_t) \eta_t$. Then
linearization methods. i.e. suitable perturbations around the optimal solution, can be applied leading to certain Riccati type equations. 
This point of view is naturally closer to the more established  semi-classical Lagrangian approach. We shall report on these approaches in more 
detail  in a separate publication \cite{DoikouMalhamWiese2}.

\begin{itemize}
\item {\tt Remark 7: Feynman-Kac formula}

Having computed the propagator explicitly (\ref{discrete1}) we conclude that equation (\ref{kernel}) can be then expressed as
\begin{equation}
f({\mathrm x}_f, t_f) =  \int d {\mathbb M}\ e^{\int_0^t u({\mathrm x})ds} f_0({\mathrm x}_0), ~~~~~f_0({\mathrm x}_0) = f({\mathrm x}_0, t_0) \nonumber
\end{equation}
which  is precisely  the Feynman-Kac formula, and describes the time evolution of a given initial profile $f_0({\mathrm x}_0)$ to $f({\mathrm x}_f, t_f) $ a solution of the Fokker-Planck equation.

One could have started from the Kolmogorov backward equation and computed the path integral backwards in time, which would have led to the time-reversed version of the Feynman-Kac formula:
\begin{equation}
f({\mathrm x}_0, t_0) =  \int d {\mathbb M}\ e^{\int_0^t u({\mathrm x})ds} f_f({\mathrm x}_f), ~~~~~f_f({\mathrm x}_f) = f({\mathrm x}_f, t_f). \nonumber
\end{equation}
In this case the Feynman-Kac formula describes the reversed time evolution of a given final state $f_f({\mathrm x}_f)$, backwards in time, to a previous state $f({\mathrm x}_0, t_0) $  
a solution of the Kolmogorov backward equation.
\end{itemize}

Let us also briefly outline the re derivation of the path integral in the case of dynamical diffusion matrices generalizing  the computations of the previous subsection, 
which led to the discrete time version of Girsanov's theorem. We start from the general expression (\ref{discrete}) and perform the Gauss integrals 
(set ${\mathbb B}_n({\mathrm x})= \Delta{\mathrm x}_n - \delta b_n({\mathrm x})$)
\begin{equation}
\int d {\bf p}_n\ \exp\Big [-{\delta \over 2} {\mathrm p}^T_n g_n({\mathrm x}) {\mathrm p}_n + {\mathrm i}{\mathrm p}_n^T {\mathbb B}_n({\mathrm x}) \Big ] =  \big (\det \sigma_n({\mathrm x}) \big)^{-1}
\Big ( {2\pi \over \delta}\Big )^{M\over 2} e^{- {1\over 2 \delta}\big (\sigma_{n}^{-1}({\mathrm x}) {\mathbb B}_n({\mathrm x})\big)^T \sigma_{n}^{-1}({\mathrm x}) {\mathbb B}_n({\mathrm x})} .
\label{gauss3} \nonumber
\end{equation}
Let us now focus on the term $\sigma_{n}^{-1}({\mathrm x}) {\mathbb B}_n({\mathrm x}) $, which can be expressed as
\begin{eqnarray}
\sigma_{n}^{-1}({\mathrm x}) {\mathbb B}_n({\mathrm x})  
= (\sigma_{n}^+)^{-1}({\mathrm x}) \Delta {\mathrm x}_n - {1\over 2}\Delta \sigma_n^{-1}({\mathrm x})\ \sigma_{n}^+({\mathrm x})\ (\sigma_{n}^+)^{-1}({\mathrm x})\Delta {\mathrm x}_n - 
\delta \sigma_{n}^{-1}({\mathrm x}) b_n({\mathrm x}),  \nonumber
\end{eqnarray}
in accordance to our usual notation: $(\sigma_n^+)^{-1} = {1\over 2}\big (\sigma_{n+1}^{-1}+\sigma_n^{-1}\big ),\ \Delta \sigma_n^{-1}= \sigma_{n+1}^{-1}-\sigma_n^{-1}$. We set
\begin{equation}
\Delta {\mathrm y}_n= (\sigma^+_n)^{-1}({\mathrm x}) \Delta {\mathrm x}_n, \label{discr3}  
\end{equation}
which is the discrete time analogue of definition (\ref{trans}). Using (\ref{discr3}) we conclude:
\begin{eqnarray}
\sigma_{n}^{-1}({\mathrm x}) {\mathbb B}_n({\mathrm x}) &=& \Delta{\mathrm y}_n - {1\over 2} \Delta \sigma_n^{-1}({\mathrm y}) \sigma_n({\mathrm y}) \Delta{\mathrm y}_n-
\delta \sigma_{n}^{-1}({\mathrm y}) b_n({\mathrm y}) \nonumber\\
&=& \Delta{\mathrm y}_n -\delta \tilde b_n({\mathrm y}) \nonumber
\end{eqnarray}
after solving ${\mathrm x}_n = {\mathrm x}_n({\mathrm y}_n)$ via (\ref{discr3}) as in the continuous time case discussed in section 2. The definition of $\tilde b$ in 
the expression above is apparently the discrete time analogue of (\ref{drifta}). Bearing also in mind the volume element change $\big (\det \sigma_n({\mathrm x})\big )^{-1} 
\prod_{j=1}^M d{\mathrm x}_{jn} =  \prod_{j=1}^M  d{\mathrm y}_{jn}$, we conclude that the propagator (\ref{discrete}) coincides with expression (\ref{exp}), 
thus all computations of subsection 3.1 apply and the findings of section 2 are once more confirmed.

\subsubsection{The case $\det \sigma =0$}

\noindent We shall briefly discuss in what follows the special situation where $\det \sigma =0$, in this case where the $\sigma$-matrix has at least one zero eigenvalue.
 Let us consider the quite general scenario, where $\sigma$  is given as
\begin{equation}
\sigma({\mathrm x})=\lb \begin{matrix}
	\bar \sigma({\mathrm x})_{m \times m}	 & 0_{(M-m)\times m}\\
   0_{m\times (M-m)}		&  0_{(M-m)\times (M- m)}
	\end{matrix} \rb. \nonumber
\end{equation}
Then the corresponding SDEs are expressed as follows
\begin{eqnarray}
&& d{\mathrm x}_{ti} = b_i({\mathrm x}_t) dt + \sum_{j=1}^m \bar \sigma_{ij}i({\mathrm x}_t) d{\mathrm w}_{tj}, ~~~~i \leq m \nonumber\\
&& d{\mathrm x}_{ti} = b_i({\mathrm x}_t) dt, ~~~~i>m. \label{st3}
\end{eqnarray}
The associated generator $\hat L_0$ is then given as
\begin{equation}
\hat L_0 = \sum_{i, j=1}^m g_{ij}({\mathrm x}) {\partial^2 \over \partial {\mathrm x}_i \partial {\mathrm x}_j} + 
\sum_{i=1}^M b_i({\mathrm x}){\partial \over \partial {\mathrm x}_i},
\end{equation}
$g =\bar \sigma \bar \sigma^T$. 

Following the findings of section 4 and assuming that $\det \sigma \neq 0$, we consider the following simple change of variables:
\begin{eqnarray}
&& d{\mathrm y}_{ti} = \sum_{i, j=1}^m \bar \sigma_{ij}^{-1}({\mathrm x}_t) d{\mathrm x}_{tj}, ~~~~~i\leq m \nonumber\\
&& d{\mathrm y}_{ti} = d{\mathrm x}_{ti}, ~~~~~i >  m, \nonumber
\end{eqnarray}
then the generator takes the simpler form
\begin{equation}
\hat L_0 = \sum_{i=1}^m {\partial^2 \over \partial {\mathrm y}^2_i } + \sum_{i=1}^m \tilde b_i({\mathrm y}) 
{\partial \over \partial {\mathrm y}_i} +  \sum_{i=m+1}^M b^-_i({\mathrm y})  {\partial \over \partial {\mathrm y}_i} \nonumber
\end{equation}
and the SDE (\ref{st3}) becomes
\begin{eqnarray}
&& d{\mathrm y}_{ti} = \tilde b_i({\mathrm y}_t) dt  + d{\mathrm w}_{tj}, ~~~~i \leq m \nonumber\\
&& d{\mathrm y}_{ti} = b^-_i({\mathrm y}_t) dt, ~~~~i > m, \label{st3b} \nonumber
\end{eqnarray}
$\tilde b$ is an $m$-vector defined as
\begin{equation}
\tilde b({\mathrm y}_s) = \bar \sigma^{-1}({\mathrm y}_s)\Big(b({\mathrm y}_s) -{ 1\over 2}  (\partial^-_{{\mathrm y}_s} \bar \sigma^T({\mathrm y}_s))^T\Big ), ~~~~~~
\partial^-_{\mathrm y} = \Big (\partial_{{\mathrm y}_1}, \ldots, \partial_{{\mathrm y}_m} \Big ), \label{drift2bb} \nonumber
\end{equation}
and $b^-$ is an $M-m$ vector with components $b_i({\mathrm y}), ~~i\in \{m+1, \ldots, M\}$. Typical systems with such a behavior are systems in the 
presence of discontinuities/defects or interfaces. The DNLS/DST model in the presence of local defects is a such a system and will be examined in the subsequent section.

\section{Lagrangian versus stochastic formulation}

\noindent
In this section we compute the propagators for certain prototype models via the Lagrangian and  stochastic approaches.
More precisely, in the next subsection we evaluate Feynman's path integral (\ref{fin0}), for a system of $M$-damped harmonic oscillators with linear drift i.e.
the case of $ M$-dimensional Ornstein-Uhlenbeck process using the standard Lagrangian point of view.
In subsection 4.2 we review the computation of the  propagator for the $1$-dimensional harmonic oscillator via the stochastic approach (see also \cite{MansuyYor}). 

We also comment  on generic processes associated to more complicated SDEs as well as on  possible future directions combining the two approaches i.e. 
Lagrangian approach and perturbation theory versus stochastic methods and use of solutions of SDEs. Here we basically consider simple prototypical 
models such as the harmonic oscillator and the
Ornstein-Uhlenbeck process, in order to review how the two approaches work. However the goal is the study of more involved examples, 
where both perturbative/semi-classical methods and the stochastic approach can be tested and compared.

\subsection{Lagrangian approach: multidimensional Ornstein-Uhlenbeck process}

\noindent
In this subsection we compute the propagator (\ref{fin0}) for a system of $M$-damped oscillators using the standard for physicists formulation i.e. the Lagrangian description. 
The quantum Hamiltonian in this case is given by (\ref{constant}) with:
\begin{eqnarray}
&& \tilde b({\mathrm y}) = - \Theta {\mathrm y}, ~~~~~u({\mathrm y}) = -{1\over 2} {\mathrm y}^T \hat \Theta^T \hat \Theta {\mathrm y}, \nonumber
\end{eqnarray}
where $\Theta,\ \hat \Theta$ are constant $M\times M$ matrices.
Notice in this case we deal with a non-self-adjoint operator. From the stochastic point of view this is a multidimensional 
Ornstein-Uhlenbeck process  (see also expression (\ref{fin0})) with relevant SDEs given by:
\begin{equation}
d{\mathrm y}_t =-\Theta{\mathrm y}_t dt + d{\mathrm w}_t.
\end{equation}

We shall express the path integral (\ref{fin0}) in a form familiar for physicists (we have suppressed the time subscript $s$ in this subsection for simplicity)
\begin{equation}
K({\mathrm y}_f, {\mathrm y}_i| t) = \int d{\bf q}\ \exp\Big [ -\int_{0}^t  {\cal L}({\mathrm y}, \dot {\mathrm y}) ds \Big ], \label{path}
\end{equation}
and the Lagrangian is given (\ref{fin0}), (\ref{1}):
\begin{equation}
{\cal L}({\mathrm y}, \dot {\mathrm y})  = {1\over 2} \dot {\mathrm y}^T \dot {\mathrm y} +{1\over 2} \tilde b^T({\mathrm y})\tilde b({\mathrm y}) -
\tilde b^T({\mathrm y})\dot{\mathrm y} +{1\over 2} \nabla_{\mathrm y}\tilde b({\mathrm y}) - u({\mathrm y}) \nonumber
\end{equation}
the third term in the expression above comes from the term $\int_0^t \tilde b^T \circ d{\mathrm y}$ in (\ref{fin02}), and $\dot {\mathrm y} = {d {\mathrm y} \over ds}$, (recall Remarks 4, 5). 
Indeed, notice that here all the involved integrals are Stratonovich type integrals, hence the extra term ${1\over 2} \nabla_{\mathrm y}\tilde b({\mathrm y}) $ (Ito's contribution (\ref{fin02})).
The astute reader might be perplexed by the appearance of $\dot {\mathrm y}$ in the expression above, however this is nothing 
but a convenient choice of notation. 

The Lagrangian can be then expressed in components as:
\begin{eqnarray}
{\cal L}({\mathrm y}, \dot {\mathrm y}) &=&  {1\over 2} \sum_i \dot {\mathrm y}_i^2  + {1\over 2}\sum_i \Omega^2_{ij} {\mathrm y}_i {\mathrm y}_j +
\sum_{i, j}\Theta_{ij} \dot {\mathrm y}_i {\mathrm y}_j  
-{\mbox{tr}\Theta \over 2}\label{ll}
\end{eqnarray}
where $\Omega^2 = \Theta^T \Theta +\hat \Theta^T \hat  \Theta $ and is by construction symmetric. Notice that in the special case where $\Theta$ 
is a symmetric matrix the third term of the expression above can 
be expressed as a total derivative giving rise to a purely boundary term after integration. 

Let us now follow the typical Lagrangian approach in computing the path integral (\ref{path}) and focus on the case where $\Theta$  is symmetric. 
Let ${\mathrm y} = {\mathrm z} + {\mathrm w}$, where ${\mathrm z}$ is the deterministic contribution i.e. 
the solution of the classical equations of motion and ${\mathrm w}$ is the random contribution, where we assume 
the following boundary conditions:  ${\mathrm z}(0) = {\mathrm z}_0,\  {\mathrm z}(t) = {\mathrm z}_f$ 
and ${\mathrm w}(0)= {\mathrm w}(t) =0$. Then we obtain
\begin{equation}
 \exp\Big [ -\int_{0}^t  {\cal L}({\mathrm y}, \dot {\mathrm y}) ds \Big ] =  \exp\Big [ -\int_{0}^t \Big ( \tilde {\cal L}({\mathrm z}, 
\dot {\mathrm z}) ds +\tilde {\cal L}({\mathrm w}, \dot {\mathrm w}) +  
 F({\mathrm z}, \dot {\mathrm z}; {\mathrm w}, \dot {\mathrm w}) \Big )ds \Big ] e^{-{1\over 2}{\mathrm z}^T_f\Theta {\mathrm z}_f  
+ {1\over 2}{\mathrm z}^T_0\Theta {\mathrm z}_0+ \mbox{tr}\Theta {t\over 2}  }  
\label{dec}
\end{equation}
where we define
\begin{equation}
\tilde {\cal L}({\mathrm z}, \dot {\mathrm z})  =  {1\over 2} \sum_i \dot {\mathrm z}_i^2  + {1 \over 2}\sum_i \Omega^2_{ij} {\mathrm z}_i {\mathrm z}_j.
\end{equation}
The classical equations of motion  are obtained via the Euler-Lagrange equations:
\begin{equation}
{\partial \tilde{\cal L} \over \partial {\mathrm z}_j} = {\partial \over \partial s}\Big 
({\partial \tilde {\cal L} \over \partial \dot {\mathrm z}_j}\Big ) \ \Rightarrow\ \ddot{\mathrm z} -  \Omega^2 {\mathrm z} =0. \label{em}
\end{equation}
The general solution of (\ref{em}) is given by
\begin{equation}
{\mathrm z}(s) =  \sinh{s\Omega } \Big ( \sinh{ t\Omega }^{-1} {\mathrm z}_f - \sinh{ t\Omega }^{-1} \cosh{t\Omega}{\mathrm z}_0 \Big ) 
+  \cosh{s \Omega}{\mathrm z}_0  \label{solution}
\end{equation}
where recall we have considered ${\mathrm z}(0) =z_0,\ {\mathrm z}(t)={\mathrm z}_f $.

The last term inside the integral in (\ref{dec}) is linear in ${\mathrm w}$, and disappears via the classical equations of motion, 
whereas the classical Lagrangian will produce only boundary terms due to (\ref{em}). 
The path integral is then expressed as
\begin{eqnarray}
&& K({\mathrm y}_f, 0| t) =  e^{-{1\over 2}{\mathrm z}_f^T \dot{\mathrm z}_f +{1\over 2}{\mathrm z}_0^T \dot{\mathrm z}_0 - 
{1 \over 2}{\mathrm z}_f^T\Theta {\mathrm z}_f + {1 \over 2}{\mathrm z}_0^T\Theta {\mathrm z}_0 
+{\Theta t \over 2} }\ \int d{\bf w} e^{-\int_0^t \tilde{\cal L}({\mathrm w}, \dot {\mathrm w}) ds} \label{path2}  \\
&&  d{\bf w}  = \lim_{\delta \to 0} \lim_{N \to \infty }\ {1 \over (2 \pi \delta )^{NM\over 2}} \  \prod_{n=2}^N \prod_{j=1}^M d{\mathrm w}_{jn}.
\nonumber
\end{eqnarray}
Let as first compute the classical contribution of the path integral (\ref{path2}) using the solution (\ref{solution}) of the classical equations of motion. 
Indeed, substituting ${\mathrm z}_f,\ \dot{\mathrm z}_f $ via (\ref{solution}) 
we immediately obtain
\begin{eqnarray}
 && \exp\Big [-{1\over 2}{\mathrm z}_f^T \dot{\mathrm z}_f +{1\over 2}{\mathrm z}_0^T \dot{\mathrm z}_0 -
{1 \over 2}{\mathrm z}_f^T\Theta{\mathrm z}_f+ {1 \over 2}{\mathrm z}_0^T\Theta{\mathrm z}_0+  \mbox{tr}\Theta {t\over 2} \Big ]= \nonumber\\
&& \exp\Big [- {1\over 2} \mbox{tr}\Big (\big ( {\mathrm z}_f {\mathrm z}_f^T +{\mathrm z}_0 {\mathrm z}_0^T\big )\cosh{t\Omega} \Omega
\sinh{ t \Omega }^{-1}\Big )+ \mbox{tr}\Big ( {\mathrm z}_0 {\mathrm z}_f^T \Omega \sinh{t\Omega}^{-1}\Big )- 
{1\over 2}\Big ({\mathrm z}^T_f \Theta{\mathrm z}_f - {\mathrm z}^T_0 \Theta {\mathrm z}_0 \Big )\Big ]e^{ \mbox{tr}\Theta {t\over 2} }. \nonumber\\ \label{classical}
\end{eqnarray}
To compute the quantum contribution of the path integral (\ref{path2}) we consider a Fourier transform expansion for the random part of the fields (see also subsection 3.2)
\begin{equation}
{\mathrm w}_s =  \sqrt{ {2\over t}} \sum_{k\geq 1} {\sin{\omega_k s}\over \omega_k} {\mathrm f}_k, ~~~~~~\omega_k = {k \pi \over t}\label{ft}
\end{equation}
${\mathrm f}_k$ are $M$-vector fields with components ${\mathrm f}_{kj}$. After some straightforward substitutions via (\ref{ft}) we obtain
\begin{eqnarray}
\int d{\bf w} e^{-\int_0^t \tilde {\cal L}({\mathrm w}, \dot {\mathrm w}) ds}
&=& {1\over (2\pi t)^{M\over 2}} \int d {\bf f}\exp \Big [- {1\over 2}\sum_{k\geq 1}{\mathrm f}_k^T \Big ({\mathbb I}_M+ {t^2 \over k^2 \pi ^2} \Omega^2\Big ){\mathrm f}_k \Big ],  \label{gauss}\\
&=&  {1\over (2\pi t)^{M\over 2}} {\mathbb E}\Big (\exp\Big [- {1\over 2}\sum_{k \geq 1}{\mathrm f}_k^T \Big ({ t^2 \over k^2 \pi ^2} \Omega^2\Big ){\mathrm f}_k\Big] \Big ) \nonumber\\
&& d {\bf f} = \prod_{k \geq 1}\prod_{j=1}^M {d{\mathrm f}_{kj} \over \sqrt{2\pi}}. \nonumber
\end{eqnarray}
As already mentioned earlier in the text the appearance of $\dot {\mathrm y}$ in the expression above although perhaps perplexing  is nothing 
but a convenient choice of notation. In any case, we are able to obtain the correct result, having suitably regularized the coefficients of the Fourier transform (\ref{ft}). 

After performing the Gaussian integrals in (\ref{gauss}) we express the quantum contribution as
\begin{equation}
\int d{\bf w} e^{-\int_0^t\tilde {\cal L}({\mathrm w}, \dot {\mathrm w}) ds}=  {1\over (2\pi t)^{M\over 2}}
\prod_{k\geq 1} \det\Big ({\mathbb I}_m +{ t^2 \over k^2 \pi ^2} \Omega^2  \Big)^{-{1\over 2}}. 
\label{quantum}
\end{equation}
Recalling  the infinite product identity
\begin{equation}
\prod_{k\geq 1} (1+{a^2\over k^2})^{-1} = {\pi a \over \sinh{\pi a}}  \nonumber
\end{equation}
we eventually obtain
\begin{equation}
\int d{\bf w} e^{-\int_0^t \tilde {\cal L}({\mathrm w}, \dot {\mathrm w}) ds}={1\over (2\pi t)^{M\over 2}} 
\det \Big ((t \Omega)^{-1}\ \sinh{t \Omega }\Big )^{-{1\over 2}}.  \label{quantum2}
\end{equation}

Putting together the classical and quantum contributions (\ref{classical}), (\ref{quantum2}) and recalling that ${\mathrm y}_{0,f} = {\mathrm z}_{0,f}$ we conclude (\ref{path2}):
\begin{eqnarray}
K({\mathrm y}_f, {\mathrm y}_0| t) &=&   { 1\over (2\pi t)^{M\over 2}}
\det \Big ((t \Omega)^{-1}\ \sinh{t \Omega }\Big)^{-{1\over 2}} \exp\Big [- {1\over 2} \mbox{tr}\Big (\big({\mathrm y}_f {\mathrm y}^T_f - {\mathrm y}_0{\mathrm y}^T_0 \big )\Theta\Big )+ 
 \mbox{tr}\Theta {t\over 2}  \Big ] \nonumber\\ &\times &
\exp\Big [- {1\over 2} \mbox{tr}\Big (\big ( {\mathrm y}_f {\mathrm y}_f^T +{\mathrm y}_0 {\mathrm y}_0^T\big )\cosh{t\Omega} \Omega
\sinh{ t \Omega }^{-1}\Big )+ \mbox{tr}\Big ( {\mathrm y}_0 {\mathrm y}_f^T \Omega \sinh{t\Omega}^{-1}\Big ) \Big ] .  \nonumber\\ \label{kernelou}
\end{eqnarray}
In the case where the potential is zero $\hat \Theta =0$, we obtain the multidimensional Ornstein-Uhlenbeck propagator. In the one dimensional 
case in particular the generic expression above reduces to 
the familiar propagator:
\begin{equation}
K({\mathrm y}, {\mathrm x}| t) = \sqrt{\Theta\over \pi (1-e^{-2\Theta t})}\ \exp\Big [{-\Theta ({\mathrm y}-{\mathrm x} e^{-\Theta t})^2 \over1-e^{-2\Theta t} }\Big ]. \label{OU1}
\end{equation}
It is clear that the Lagrangian description provides a straightforward way of computing the propagator of the Ornstein-Uhlenbeck process. 

Let us comment on the case where $\Theta$ contains an anti symmetric part as well, i.e. $\Theta = \Sigma + A$ where $\Sigma$ is the symmetric part and $A$ is the anti symmetric one.
From a physical viewpoint this case corresponds to the presence of an angular momentum term in the Lagrangian, whereas from the computational point of view there are two non-trivial 
contributions when computing the propagator.  First the classical equations of motion are modified due to the existence of the anti-symmetric part. 
More precisely, from the Euler-Lagrange equations one obtains, the following classical equations of motion:
\begin{equation}
\ddot{\mathrm z} +2 A\dot{\mathrm z} - \Omega^2 {\mathrm z} =0, \nonumber
\end{equation}
thus the classical contribution (\ref{classical}) is also modified accordingly. More importantly there is a highly non-trivial contribution when computing 
the quantum part (\ref{ft}), indeed  when $\Theta$ is symmetric as described above the term 
$e^{-\int_0^t ds \sum_{i, j}\Theta_{ij} (\dot {\mathrm w}_i {\mathrm w}_j- {\mathrm w}_i\dot {\mathrm w}_j)}$ gives a purely boundary contribution. However, 
when there is an anti-symmetric part then an extra non-trivial term in the integral (\ref{ft}) appears, which should be taken into consideration. 
The computation of the quantum contribution when $\Theta$ is anti-symmetric and for $\Omega =0$  is essentially equivalent to the problem 
of computing the characteristic function for the Levy area:
\begin{equation}
{\mathbb A }= \int d{\bf m}_0 \exp\Big [ \sum_{i>j} \Theta_{ij} \int({\mathrm w}_{si} d{\mathrm w}_{sj} - {\mathrm w}_{sj}d{\mathrm w}_{si} )\Big ] \label{Levy}
\end{equation}
where $d{\bf m}_0$ is given in (\ref{measure1}), and ${\mathrm w}_s$ is given in (\ref{random}) (see also e.g. \cite{Wiktorsson} and references therein). 
This is a significant issue, which however will be addressed in a separate publication \cite{DoikouMalhamWiese2}.

\subsection{Stochastic approach: one dimensional harmonic oscillator}

\noindent 
Our goal in this subsection is to rederive the propagator for the one dimensional quantum harmonic oscillator using probabilistic techniques, 
although the analytical form of the Feynman's path integral via the Lagrangian (semi-classical) formulation is widely known in the quantum physics community, 
given by the celebrated Mehler formula. 
Consider the single particle quantum harmonic oscillator, with potential $u({\mathrm x})={\omega^2\over 2}{\mathrm x}^2$ for ${\mathrm x} \in\R$, where 
$\omega>0$ is a fixed parameter.
We derive the formula here from the probabilistic perspective. Using  (\ref{B}), (\ref{measure1}), we evaluate:
\begin{eqnarray}
K({\mathrm y},{\mathrm x}|t) &= &\int d{\mathbb M} \ \exp\biggl(-\tfrac12\omega^2\int_0^t|{\mathrm x}_s|^2\,\rd s\biggr) \nonumber\\
&=&{e^{-{({\mathrm x}-{\mathrm y})^2 \over 2t}}\over \sqrt{2\pi t}}\mathbb E\, 
\Biggl(\exp\biggl(-\tfrac12\omega^2\int_0^t|{\mathrm x}_s|^2\,\rd s\biggr)\colon {\mathrm x}_t={\mathrm y}, {\mathrm x}_0={\mathrm x}\Biggr ). 
\label{basic2} \nonumber
\end{eqnarray}
For ${\mathrm x}=0$ this is a standard quadratic functional of Brownian motion studied
extensively by Pitman and Yor, see for example 
\cite{MansuyYor}.

We now derive this result for ${\mathrm x}\neq 0$. Let ${\mathrm w}_s$ be 
a standard Wiener process (${\mathrm w}_0 =0$). Then the required
Brownian bridge satisfying ${\mathrm x}_0={\mathrm x}$ and ${\mathrm x}_t={\mathrm y}$ is given by (\ref{sequation})
${\mathrm w}_t = {\mathrm y}-{\mathrm x}$ and as Wiener proved a standard Wiener process has the
Fourier representation on $[0,\ t]$  given in (\ref{random}), where  ${\mathrm f}_n$ are normal random variables, 
(${\mathrm w}_t=\sqrt{t}\,{\mathrm f}_0$). 
Substituting this Fourier series into the prescription for ${\mathrm x}_s$ above
\begin{equation}
{\mathrm x}_s={\mathrm x}+\frac{s}{t}({\mathrm y}-{\mathrm x})+\frac{\sqrt{2t}}{\pi}
\sum_{n\geq1}\frac{1}{n}{\mathrm f}_n\,\sin{\frac{n\pi s}{t}}.
\end{equation}

Let us directly substitute the Fourier series for ${\mathrm x}_s$ into 
the integral in the exponent. After performing standard trigonometric integrals we conclude
\begin{equation*} 
-\tfrac12\omega^2\int_0^t|{\mathrm x}_s|^2\,\rd s
=-\tfrac12\sum_{n\geqslant1} \bigl(a_n{\mathrm f}_n^2+b_n{\mathrm f}_n\bigr)-\tfrac12 c,
\end{equation*}
where we define
\begin{equation*} 
a_n\colon =\frac{\omega^2t^2}{\pi^2n^2},\quad
b_n\colon =\frac{\omega^2(2t)^{3/2}}{\pi^2n^2}\bigl({\mathrm x}-(-1)^n{\mathrm y}\bigr)
\quad\text{and}\quad
c\colon =\frac{\omega^2t}{3}({\mathrm x}^2+{\mathrm x}{\mathrm y}+{\mathrm y}^2).
\end{equation*}

The expectation value for a generic term in the sum, namely\\
$\mathbb E\, \Biggl(\exp\bigl(-\frac12 a_n\xi^2-\frac12 b_n\xi\bigr)\Biggr)$, with $\xi$ a standard normal
random variable. Directly computing we find  
\begin{align*} 
\mathbb E\,\,\Biggl(  \exp\bigl(&-\tfrac12 a_n\xi^2-\tfrac12 b_n\xi\bigr)\,\Biggr )\\
=&\;\frac{1}{\sqrt{2\pi}}\exp\Bigl(\tfrac12\frac{\omega^4b_n^2}{4(1+\omega^2a_n)}\Bigr)
\int_{\R}\exp\Bigl(-\tfrac12(1+\omega^2a_n)(\xi+C)^2\Bigr)\,\rd\xi\\
=&\;\exp\Bigl(\tfrac12\frac{\omega^4b_n^2}{4(1+\omega^2a_n)}\Bigr)\cdot\frac{1}{(1+\omega^2a_n)^{1/2}},
\end{align*}
where $C=\omega^2b_n/\bigl(2(1+\omega^2a_n)\bigr)$.

Substituting these last expressions into the expectation of interest, we find  
\begin{align*} 
\mathbb E\,&\,\Biggl(\exp\biggl(-\tfrac12\omega^2\int_0^t|{\mathrm x}_s|^2\,\rd s\biggr)\,\Biggr)\\
=&\;\exp\Bigl(-\frac{\omega^2t}{6}({\mathrm x}^2+{\mathrm x}{\mathrm y}+{\mathrm y}^2)\Bigr)
\Biggl(\prod_{n=1}^\infty\frac{n^2}{n^2+\omega^2t^2/\pi^2}\Biggr)^{1/2} \exp\Biggl(\frac{(2t)^3}{8}\frac{\omega^4}{\pi^4}
\sum_{n\geqslant1}\frac{n^{-4}\bigl({\mathrm x}-(-1)^n{\mathrm y}\bigr)^2}{1+\omega^2t^2/(n^2\pi^2)}\Biggr)\\
=&\;\exp\Bigl(-\frac{\omega^2t}{6}({\mathrm x}^2+{\mathrm x}{\mathrm y}+{\mathrm y}^2)\Bigr)
\Biggl(\frac{\omega t}{\mathrm{sinh}\,(\omega t)}\Biggr)^{1/2} \exp\Biggl(t^3\frac{\omega^4}{\pi^4}
\sum_{n\geqslant1}\frac{{\mathrm x}^2+{\mathrm y}^2-2(-1)^n{\mathrm x}{\mathrm y}}{n^2(n^2+\omega^2t^2/\pi^2)}\Biggr),
\end{align*}
where we have substituted for $a_n$ and $b_n$ from the expressions above
and also used Euler's formula for the infinite product shown.

We now focus on the terms in the exponent of the final factor. 
By direct computation we observe that 
\begin{align*} 
t^3\frac{\omega^4}{\pi^4}
\sum_{n\geqslant1}&\frac{{\mathrm x}^2+{\mathrm y}^2-2(-1)^n{\mathrm x}{\mathrm y}}{n^2(n^2+\omega^2t^2/\pi^2)}\\
=&\;t\frac{\omega^2}{\pi^2}\Biggl(\bigl({\mathrm x}^2+{\mathrm y}^2\bigr)
\biggl(\frac{\pi^2}{6}+\frac{\pi^2}{\omega^2t^2}\tfrac12\bigl(1-\omega t\,\mathrm{coth}\,(\omega t)\bigr)+2{\mathrm x}{\mathrm y}
\biggl(\frac{\pi^2}{6}+\frac{\pi^2}{\omega^2t^2}
\tfrac12\bigl(1-\omega t\,\mathrm{coth}\,(\omega t)\bigr)\biggr)\\
&\;-{\mathrm x}{\mathrm y}\biggl(\frac{\pi^2}{6}+\frac{4\pi^2}{\omega^2t^2}
\tfrac12\bigl(1-\tfrac12\omega t\,\mathrm{coth}\,(\tfrac12\omega t)\bigr)\biggr)\Biggr).
\end{align*}

Suitably combining the above contributions we conclude that the quadratic functional of the Brownian bridge associated with the 
scalar single particle quantum harmonic oscillator can be then expressed as follows 
\begin{multline*}
\mathbb E\,\Biggl(\exp\biggl(-\tfrac12\omega^2\int_0^t|{\mathrm x}_s|^2\,\rd s\biggr)\colon {\mathrm x}_t={\mathrm y}, {\mathrm x}_0={\mathrm x} \Biggr)\\
=\Biggl(\frac{\omega t}{\mathrm{sinh}\,(\omega t)}\Biggr)^{1/2}
\exp\Biggl(-\frac{\omega}{2}\bigl({\mathrm x}^2+{\mathrm y}^2\bigr)\mathrm{coth}\,(\omega t)
+\frac{\omega{\mathrm x}{\mathrm y}}{\mathrm{sinh}\,(\omega t)} + {({\mathrm x}-{\mathrm y})^2\over 2t}\Biggr).
\end{multline*}
which gives
\begin{equation*}
K({\mathrm y},{\mathrm x}|t) = \Biggl(\frac{\omega }{2\pi \mathrm{sinh}\,(\omega t)}\Biggr)^{1/2}
\exp\Biggl(-\frac{\omega}{2}\bigl({\mathrm x}^2+{\mathrm y}^2\bigr)\mathrm{coth}\,(\omega t)
+\frac{\omega{\mathrm x}{\mathrm y}}{\mathrm{sinh}\,(\omega t)}\Biggr).
\end{equation*}
and this is precisely Mehler's formula. We observe that the proof above reveals that the physicist's proof of Mehler's formula
via the action $S[{\mathrm x}]$ is much more succinct. Note that the extension of Mehler's formula for the $M$-particle 
harmonic oscillator by means of probabilistic  techniques is given in \cite{DoikouMalhamWiese2}, but also the results of 
the previous section from the semi-classical point of view are relevant when considering the case of zero drift.

\subsubsection{Examples of stochastic processes $\&$ applications}

\noindent 
In general, one of the main aims is the computation of expectation values using the universal expression 
(\ref{discrete1}), (\ref{measure1}) and solutions of the underlying SDEs. In particular, in statistical and quantum physics 
the quantities of significance are (we use here the familiar in physics community  notation for expectation values $\langle {\cal O} \rangle$):
\begin{eqnarray}
\langle {\cal O}({\mathrm x}_s ) \rangle ={{\mathbb  E}_t\Big ({\cal O}({\mathrm x}_s )\ e^{\int_0^t u({\mathrm x}_s) ds} \Big )\over 
{\mathbb E}_t\Big (e^{\int_0^t u({\mathrm x}_s) ds} \Big )}, 
~~~~0 \leq s \leq t \label{stochI}
\end{eqnarray}
where we define via (\ref{discrete1}), (\ref{measure1})
\begin{eqnarray}
&& {\mathbb E}_t\Big ({\cal O}({\mathrm x}_s)\Big )=\int  d{\bf w_t}\ d{\mathbb M}\ {\cal O}({\mathrm x}_s) ~~~~0 \leq s \leq t . \label{stochII} 
\end{eqnarray}
Formula (\ref{stochI}) can be practically used provided that solutions of the associated SDEs (SPDEs in the continuum space limit, i.e. in quantum/statistical field theories) 
are available so that the fields ${\mathrm x}_{tj}$  are expressed in terms of the normal variables ${\mathrm w}_{tj}$.   Alternatively, in the cases of generic drift and/or 
potential in order to obtain the propagator and expectation values one can in principle start from (\ref{fin0}) and apply semi-classical methods, perturbation theory, and  
Green's function techniques as is customary in quantum and statistical field theory.

Let us  briefly discuss below a few illustrative examples of simple processes, in the absence of potential $u$, which will also be used later in the text when discussing two 
fundamental integrable quantum systems, i.e. the DST and  XXZ models. In the examples below we use solutions of the underlying SDEs in order to compute expectation values. 
\begin{enumerate}

\item {\tt Multidimensional Brownian motion with constant diffusion matrix} 

The corresponding SDEs and solutions are given by
\begin{equation}
d{\mathrm y}_t =  \sigma  d{\mathrm w}_t\  \Rightarrow \ {\mathrm y}_{sj} = {\mathrm y}_{0j} + \sum_{j}\sigma_{ij}{\mathrm w}_{sk} \label{eq1}
\end{equation}
where $\sigma$ is a non-dynamical (i.e. does not depend on the fields ${\mathrm y}_j$) $M\times M$ matrix. Let us compute the first couple of expectation values (first couple of moments): 
${\mathbb E}_t({\mathrm y}_{sj})$ and ${\mathbb E}_t({\mathrm y}_{si}\ {\mathrm y}_{s' j}) $.  Indeed, one readily finds via (\ref{stochII})
\begin{eqnarray}
&& {\mathbb E}_t({\mathrm y}_{sj}) ={\mathrm y}_{0j},  \label{a1} \nonumber\\
&& {\mathbb E}_t({\mathrm y}_{si}\ {\mathrm y}_{s' j}) =  {\mathbb E}_t({\mathrm y}_{0i}\ {\mathrm y}_{0 j}) +  {\mathbb E}_t(\sum_{k,l } \sigma_{ik} \sigma_{jl} {\mathrm w}_{sk}\ 
{\mathrm w}_{s' l}) = {\mathrm y}_{0i} {\mathrm y}_{0j}+  g_{ij} s', ~~~s' \leq s  \nonumber
\end{eqnarray}
and we have used the fact that 
\begin{eqnarray}
&& {\mathbb E}_t({\mathrm w}_{sj}) =0,  \label{a2} \\
&& {\mathbb E}_t({\mathrm w}_{si}\ {\mathrm w}_{s' j}) = \delta_{ij}s'. ~~~~s' \leq s. \label{b2}
\end{eqnarray}
Equation (\ref{a2}) is obvious via (\ref{stochII}), an explicit proof of (\ref{b2}) is presented below.

We focus  on the one dimensional case, but the multidimensional generalization is straightforward. 
This expectation value can be readily  computed using the fact that ${\mathrm w}_s- {\mathrm w}_{s'}$ and 
${\mathrm w}_{s'}$ are independent increments, which leads to:
\begin{equation}
{\mathbb E}_t({\mathrm w}_s {\mathrm w}_{s'}) = {\mathbb E}_t({\mathrm w}^2_{s'}) =s',  
~~~~s'\leq s. \label{indep} 
\end{equation}
However, let us show explicitly (\ref{indep}), via the use of  (\ref{stochII}) and Wiener's representation
 (\ref{random}).  Let $0\leq s' \leq s\leq t$:
\begin{eqnarray} 
&& {\mathbb E}_t({\mathrm w}_s {\mathrm w}_{s'})  =\nonumber\\ 
&& \int { d{\mathrm w}_t \over \sqrt{2 \pi t}} \prod_{k \geq 1}{d{\mathrm f}_k\over \sqrt{2\pi}} \exp \Big [-{{\mathrm w}_t^2 \over 2t }-\sum_{k\geq 1} {{\mathrm f}_k^2 \over 2} \Big ]
\Big ( {{\mathrm w}_t^2\over t^2}s s' + {2\over t} \sum_{k\geq 1} {\sin {\omega_k s} \sin{\omega_k s'} \over \omega_k^2}{\mathrm f}_k^2 \Big )= \nonumber\\
&&  {ss' \over t} - t \sum_{k \geq 1} {\cos{2k \pi{s+ s'\over 2t}} - \cos{2k \pi{s -s'\over 2t}} \over k^2 \pi^2}. \nonumber
\end{eqnarray} 
Taking into consideration the identity 
\begin{equation}
\sum_{k \geq 1}{\cos{2k \pi x} \over k^2 \pi^2 } = B_{2}(x), ~~~~0\leq x \leq 1, \nonumber
\end{equation}
where $B_2$ is the second Bernoulli polynomial $B_2(x) = x^2 -x +{1 \over 6}$, we conclude
\begin{equation}
{\mathbb E}_t({\mathrm w}_s {\mathrm w}_{s'})  ={ss' \over t} -t\Big (B_2({s+ s'\over 2t}) - B_2({s-s'\over 2t})\Big )= s', \nonumber
\end{equation}
which immediately leads to ${\mathbb E}_t\Big( ({\mathrm w}_s- {\mathrm w}_{s'})^2\Big ) = s -s'$.


\item {\tt Multi-dimensional geometric Brownian motion (Black-Scholes model)}

In this case  the SDEs read as
\begin{equation}
d{\mathrm x}_{tj} = b_j {\mathrm x}_{tj} dt + {\mathrm x}_{tj} \sum_k S_{jk} d{\mathrm w}_{tk} \label{BS}
\end{equation}
both $b_j$ and $S_{ij}$ are non-dynamical quantities. The obvious change of variable ${\mathrm x}_i = e^{{\mathrm y}_j}$, also in the spirit of 
the quantum canonical transform, leads to the simplified version of the equations above
\begin{equation}
d {\mathrm y}_{tj} = \tilde b_j dt + \sum_{k} S_{jk} d{\mathrm w}_{tk}\ \Rightarrow {\mathrm y}_{sj} ={\mathrm y}_{0j} + \tilde b_j s + \sum_{k} S_{ik} {\mathrm w}_{sk} \label{solutiongb}
\end{equation}
where $\tilde b_j = b_j - {g_{jj}\over 2}$. It is then straightforward to compute expectation values via the solutions  (\ref{solutiongb}) and (\ref{a2}), (\ref{b2})
\begin{equation}
{\mathbb E}_t({\mathrm x}_{sj}) = {\mathbb E}_t(e^{{\mathrm y}_{sj}}) = {\mathrm x}_{0j} e^{b_j s}, \nonumber
\end{equation}
and similarly, using the definition (\ref{stochII}) and the change of variables we obtain:
\begin{equation}
{\mathbb E}_t({\mathrm x}_{ti}{\mathrm x}_{tj}) = {\mathrm x}_{0i}{\mathrm x}_{0j} e^{(b_i + b_j)t} e^{g_{ij} t}. \nonumber
\end{equation}

As in the previous example we find the propagator via the universal expressions 
(\ref{discrete1}), (\ref{measure1}) and (\ref{solutiongb}). From the solution of  (\ref{solutiongb})
\begin{equation}
{\mathrm w}_t = S^{-1}\big (\Delta{\mathrm y}-\tilde b t\big ) \nonumber
\end{equation}
here $\Delta{\mathrm y}= {\mathrm y}_t -{\mathrm y}_0$ .
Then substituting the above in  (\ref{measure1})  we conclude that the propagator is (\ref{discrete1})
\begin{equation}
K({\mathrm y}_t, {\mathrm y}_0|t)  = {e^{- {1\over 2t}(\Delta{\mathrm y}-\tilde b t)^T g^{-1}
 (\Delta {\mathrm y}-\tilde b t)} \over {\det S}\  (2\pi t)^{M\over 2}}. \nonumber
\end{equation}
The propagator can also be expressed in terms of the vector field ${\mathrm x}_t$, recall ${\mathrm y}_j = \log {\mathrm x}_j$.

\item {\tt  Feynman-Kac $\&$ quantum quenches }

\noindent We illustrate below how the Feynman-Kac formula (\ref{kernel0}), (\ref{B}) can be utilized for the study of quantum quenches. 
We recall that a quantum quench is a process  where a quantum system is arranged in an initial state, which is an eigenstate of a Hamiltonian $ H_0 $,
and then the system evolves  in time under a different Hamiltonian $ H= H_0+H_1$.

The  Feynman-Kac formula provides indeed the time evolution of an initial state $f_0({\mathrm x}) =f({\mathrm x}, 0)$ at $t=0$ to a state $f({\mathrm y}, t)$ 
under the Hamiltonian $H=\hat L$ as discussed in section 3 . 
We consider here as an example the simple Hamiltonian $H$ of $M$-harmonic oscillators,
and as an initial profile function i.e. initial quantum state the ground state of the  Hamiltonian 
$H_0 = {1\over 2}\sum_{j}^m \partial_{{\mathrm x}_j}^2 - \sum_{j=1}^{m}{\Omega_j^2 \over 2}{\mathrm x}^2_j$ 
that describes $m<M$ harmonic oscillators:
\begin{equation}
f_0({\mathrm x}) = \exp \Biggl (-{1\over 2}\sum_{j=1}^m \Omega_j{\mathrm x}_j^2 \Biggr ) \label{f0}
\end{equation}
with  corresponding eigenvalue $E_0^{(m)} = \sum_{j=1}^m {\Omega_j \over 2}$.

The time propagator $K$ associated to the Hamiltonian of $M$-harmonic oscillators is known and is given by the generalized Mehler formula.
In fact, this formula can be easily obtained from expression (\ref{kernelou}), 
after setting ${\mathrm y}_f = {\mathrm y},\ {\mathrm y}_0 = {\mathrm x}$, $\Theta =0$ and considering  for simplicity $\Omega$ to be diagonal, we conclude 
that the propagator for the system of $M$-harmonic oscillators is given as:
\begin{equation}
K({\mathrm y}, {\mathrm x}|t) = {1\over (2\pi)^{M\over 2} } \prod_{j=1}^M  \Biggl ({\Omega_j \over \sinh{t\Omega_j}} \Biggr )^{1\over 2}\ 
\exp \Biggl ( -{1\over 2} \sum_{j=1}^M \big ( {\mathrm x}_j^2 + {\mathrm y}_{j}^2 \big ){\Omega_j \cosh{t\Omega_j} \over \sinh{t\Omega_j}}+
\sum_{j=1}^M  {\mathrm x}_j {\mathrm y}_j {\Omega_j \over \sinh{t\Omega_j} } \label{kk}
\Biggr ).
\end{equation}

Substituting $K$ (\ref{kk})  
and $f_0$ (\ref{f0}) in (\ref{kernel0}) and performing the Gaussian integrals involved we conclude
\begin{equation}
f({\mathrm y}, t)=\prod_{j=1}^m e^{-{t\Omega_j\over 2}}  \prod_{j=m+1}^M  \Biggr  ({1\over \cosh{t\Omega_j}} \Biggr )^{1\over 2}\ 
\exp\Biggl (-{1\over 2}\sum_{j=1}^m \Omega_j {\mathrm y}_j^2 -{1\over 2}\sum_{j=m+1}^M \Omega_j \coth{t \Omega_j} {\mathrm y}_j^2 \Biggr ).
\end{equation}
Consider now the behavior of the system after long enough time $t\gg 1$, i.e. when the system reaches equilibrium, then the state $f({\mathrm y}, t)$ is 
given by the ground state of the Hamiltonian of $M$-harmonic oscillators
\begin{equation}
f({\mathrm y}, t) \sim e^{-tE_0^{(M)}} \exp\Big ( -{1\over 2}\sum_{j=1}^M\Omega_j {\mathrm y}_j^2\Big ).
\end{equation}

Of course the system we examine here is simple and the underlying SDE is easy to solve. When considering more complicated diffusion reaction systems,
as will be discussed in the next section, the Feynman-Kac can be utilized provided that solutions of the SDE are available.
The time evolution of more involved systems will be discussed in future works (see e.g. \cite{DoikouMalhamWiese2}).

\end{enumerate}

\section{Discrete integrable  quantum systems $\&$ SDEs}

\noindent In this section we explore some interesting links between discrete quantum systems and SDEs. The correspondence 
between the quantum equations of motion and the SDEs is discussed and typical examples
of exactly solvable quantum systems and the associated SDEs are presented.

\noindent To illustrate the main ideas of the quantum canonical transformation and the quantum equations of motion in connection 
with SDEs we present below typical examples of integrable quantum systems:
the discrete NLS model and the Heisenberg (XXZ) quantum spin chain. As in the classical case integrability at the quantum level can 
also be described via the Lax pair $\Big ({\mathbb L}_m,\ {\mathbb A}_m\Big )$ 
which depends on quantum fields and a spectral parameter in general. The Lax pair satisfies the quantum auxiliary problem 
and hence the zero curvature condition
\begin{equation}
{d{\mathbb L}_m(\lambda)\over dt}= {\mathbb A}_{m+1}(\lambda){\mathbb L}_m(\lambda)-{
\mathbb L}_m(\lambda){\mathbb A}_{m}(\lambda) \nonumber
\end{equation}
providing the quantum equations of motion in analogy to Heisenberg's picture \cite{Korepin, DoikouFindlay}. 
The Lax pair formulation ensures the existence of many conserved charges. Indeed, define
\begin{equation}
{\mathfrak t}(\lambda) = tr_0T_0(\lambda) \nonumber
\end{equation}
where $T$ is the monodromy -a discrete path integral--is given as
\begin{equation}
T_0(\lambda) ={\mathbb L}_{0M}(\lambda)\ {\mathbb L}_{0M-1}(\lambda) \ldots {\mathbb L}_{01}(\lambda). \label{mono}
\end{equation}
Using the quantum zero curvature condition and assuming periodic boundary conditions it is shown in straightforward manner that:
${d{\mathfrak t}(\lambda) \over dt} = 0$, i.e. ${\mathfrak t}$ is the generator of the conserved quantities: ${\mathfrak t}(\lambda) = 
\sum_{k} {I_k \over \lambda^k}$

Let us clarify the index notation in the expressions above, the indices $0$ (auxiliary) and $m \in\{1, \ldots, M\}$ (quantum) correspond to 
the position in the $M+1$ tensor product
\begin{equation}
{\mathbb L}_{0m}(\lambda) =  \sum_{a,b} e_{ab}^{(0)} \otimes {\mathbb I}^{(1)} \ldots \otimes {\mathbb I}^{(m-1)} 
\otimes l^{(m)}_{ab}(\lambda) \otimes \ldots \otimes{\mathbb I}^{(M)} \nonumber
\end{equation}
where $e_{ab}$ are in general ${\cal N}\times {\cal N}$ matrices with elements: $(e_{ab})_{cd} = \delta_{ac} \delta_{bd}$. 
Notice that the quantum indices are suppressed in the expression for the monodromy (\ref{mono}) for brevity.

The algebraic formulation of quantum integrability on the other hand is based on the existence of a quantum $R$-matrix
 that satisfies the Yang-Baxter equation \cite{Korepin, Baxter}. 
This provides a rather stronger frame in the sense 
that guarantees the existence of many charges in involution, i.e. ensures the existence of a family of mutually 
commuting quantum charges: $[{\mathfrak t}(\lambda),\ {\mathfrak t}(\lambda')] =0$. 
In this setting the ${\mathbb L}$ 
operator satisfies the fundamental algebraic relation \cite{Faddeev, Korepin},
\begin{equation}
R(\lambda_1 -\lambda_2)\ \big ({\mathbb L}(\lambda_1) \otimes {\mathbb I} \big )\   \big  ({\mathbb I} \otimes {\mathbb L}(\lambda_2) \big  )=
 \big  ({\mathbb I} \otimes {\mathbb L}(\lambda_2)  \big )\  \big  ({\mathbb L}(\lambda_1) \otimes {\mathbb I}  \big )\ 
R(\lambda_1 -\lambda_2)\  \label{algebra}
\end{equation}
$R(\lambda) \in \mbox{End}(V\otimes V)$ and ${\mathbb L}(\lambda)\in \mbox{End}(V) \otimes {\cal A}$; ${\cal A}$ 
is the underlying deformed algebra defined by (\ref{algebra}). More specifically, equation 
(expressed in the index free notation) (\ref{algebra}) acts on $V \otimes V \otimes {\cal A}$
\begin{eqnarray}
&& R(\lambda) = \sum_{a,b,c,d} R_{ab, cd}(\lambda) e_{ab} \otimes e_{cd} \otimes {\mathbb I} \nonumber\\
&& {\mathbb I} \otimes {\mathbb L}(\lambda) = \sum_{i,j} e_{ab} \otimes {\mathbb I}  \otimes l_{ab}(\lambda) \nonumber\\
&& {\mathbb L}(\lambda) \otimes  {\mathbb I}=  \sum_{a,b} {\mathbb I}\otimes  e_{ab}  \otimes l_{ab}(\lambda), \nonumber
\end{eqnarray}
$l_{ij}(\lambda) \in {\cal A}$ and $T(\lambda) \in \mbox{End}(V) \otimes {\cal A}^{\otimes M}$, i.e. the monodromy is a tensor representation of (\ref{algebra}). 
In both examples we are considering here the associated $R$-matrix is the Yangian or the trigonometric XXZ $R$ matrix (see \cite{Faddeev, Korepin} and references therein).

\subsection{The quantum discrete NLS hierarchy}

\noindent 
We start our analysis with the DNLS model, with the corresponding Lax operator given by \cite{Sklyanin, KunduRagnisco}
\begin{eqnarray}
{\mathbb L}_j(\lambda) =
\begin{pmatrix}
  \lambda + {\mathbb N}_j &z_j\cr
 Z_j & 1
\end{pmatrix}. \nonumber
\end{eqnarray}
${\mathbb N}_j =\Theta_j + z_jZ_j $.
Using the fact that ${\mathbb L}$ satisfies (\ref{algebra}) together with the quantum zero curvature condition one can also construct the ${\mathbb A}$ 
operators for the whole hierarchy \cite{DoikouFindlay}. In particular, the ${\mathbb A}$-operators associated to the second and third conserved charges of the hierarchy is
\begin{eqnarray}
{\mathbb A}^{(1)}_j(\lambda) = \begin{pmatrix}
  \lambda  &z_j\cr
 Z_{j-1} & 0
\end{pmatrix},  ~~~~{\mathbb A}^{(2)}_j(\lambda) = \begin{pmatrix}
  \lambda^2 -z_j Z_{j-1}  &\lambda z_j -z_j {\mathbb N}_j +z_{j+1}\cr
 \lambda Z_{j-1}-Z_{j-1}{\mathbb N}_{j-1} +Z_{j-2} & z_j Z_{j-1}
\end{pmatrix}. \label{Aop}
\end{eqnarray}
Notice that the first charge gives  the numbers of particles of the DNLS model (corresponds to a system of $M$ harmonic oscillators), 
the second charge corresponds to the momentum of the model also known as the DST model \cite{Eilbeck}, whereas the third charge is the DNLS 
Hamiltonian \cite{KunduRagnisco}.  We shall examine below the second and third charges, identify 
the corresponding SDEs, and study their continuum limits, which produce the stochastic transport and stochastic heat equation correspondingly.

\subsubsection*{The DST model}
\noindent
Let us first focus  on the second charge (momentum) is given by \cite{Eilbeck, Sklyanin}
\begin{equation}
H^{(1)} = {1\over 2} \sum_{j=1}^M z_j^2 Z_j^2 + \sum_{j=1}^M\big (c_j z_j - z_{j+1}\big )Z_j \label{DST0} \nonumber
\end{equation}
and via (\ref{algebra}) one obtains: $[z_i,\ Z_j]= -\delta_{ij}$.
The equations of motion can now be derived via the zero curvature condition or Heisenberg's equation and they read as
\begin{equation}
{dz_j \over dt}  = \big (c_j z_j - z_{j+1}\big ) + z^2_jZ_j.\label{qem}
\end{equation}

Consider now the following map
\begin{equation}
z_ j\mapsto {\mathrm x}_j,~~~~~Z_j \mapsto \partial_{{\mathrm x}_j} \label{map1}
\end{equation}
then the Hamiltonian is expressed as:
\begin{equation}
H^{(1)} = {1\over 2} \sum_{j=1}^M {\mathrm x}_j^2 \partial_{{\mathrm x}_j}^2 + 
\sum_{j=1}^M\big (c_j {\mathrm x}_j - {\mathrm x}_{j+1}\big )\partial x_j \label{DST} \nonumber
\end{equation}
and we have considered periodic boundary conditions ${\mathrm x}_j = {\mathrm x}_{j+M}$.
The Hamiltonian is apparently of the form (\ref{general}), and the corresponding set of SDEs are given by
\begin{equation}
d{\mathrm x}_{tj}  = \big (c_j {\mathrm x}_{tj} - {\mathrm x}_{tj+1}\big )dt + {\mathrm x}_{tj} d{\mathrm w}_{tj}. \label{SDE1}
\end{equation}
If we compare the SDEs (\ref{SDE1}) with the quantum equations of motion (\ref{qem}) we observe that the last term in (\ref{qem}) 
is replaced by the multiplicative noise in (\ref{SDE1}). It is worth noting that in the special case that $c_j\gg1$ the second term in 
the drift is neglected and the SDEs (\ref{SDE1}) reduce to the ones of the multidimensional Black-Scholes model (\ref{BS}).

Let us apply the quantum canonical transformation as described in section 2. Indeed, we define the new variables ${\mathrm y}_j$ via (\ref{frame})
\begin{equation}
d {\mathrm y}_i = {\mathrm x}_i^{-1} d{\mathrm x}_i\  \Rightarrow\  {\mathrm x}_i =A_i e^{{\mathrm y}_i} \nonumber
\end{equation}
where $A_i$ are integration constants. Then the Hamiltonian of the DST model can be re-expressed in terms of the new variables as a Hamiltonian with identity diffusion matrix
\begin{equation}
H^{(1)}= {1\over 2} \sum_{j=1}^M \partial_{{\mathrm y}_j}^2 +  \sum_{j=1}^M \big  (C_j  + B_j e^{{\mathrm y}_{j+1} - {\mathrm y}_j}\big )\partial _{{\mathrm y}_j}, \label{DST2}
\end{equation}
$C_j = c_j -{1\over 2},\ B_j = -A_{j+1} A_j^{-1}$, and the corresponding set of SDEs are
\begin{equation}
d{\mathrm y}_{tj} =\big (C_j + B_j e^{{\mathrm y}_{tj+1} - {\mathrm y}_{tj}}\big )dt + d{\mathrm w}_{tj}. \nonumber
\end{equation}
Connection with the quantum Darboux transforms introduced in \cite{DoikouFindlay} would also be a very 
interesting direction to pursue. Note that in \cite{Korff, DoikouFindlay} an alternative version of the quantum discrete NLS  model is studied i.e. the so called $q$-Boson or quantum Ablowitz-Ladik model.
Recall that in the classical case the Darboux-B\"{a}cklund transformation \cite{ZakharovShabat, MatveevSalle} provides an efficient way to 
find solutions of integrable non-linear PDEs. In particular, the transformation connects solutions of the same or different integrable PDEs. 
At the quantum level the transformation connects different realizations of the underlying quantum algebra. The pertinent question is how this transformation affects the associated SDEs. 
For instance does the quantum Darboux transformation connect solutions of different SDEs? This is a significant open question, which needs to be systematically addressed.  
In fact, a novel generalization of the ``dressing'' method has been introduced for classical PDEs \cite{BeckDoikouMalhamStylianidis}, whereas a similar 
formulation is suitably extended in the context of SPDEs \cite{DoikouMalhamWiese}. For a relevant discussion on stochastic B\"{a}cklund transformation see also \cite{oconnell}.

An interesting observation can be made on the connection of the DST model and the Toda chain. Let us also consider the adjoint operator of (\ref{DST2}), which reads as
\begin{equation}
H^{(1)\dag}= {1\over 2} \sum_{j=1}^M \partial_{{\mathrm y}_j}^2 -  \sum_{j=1}^M \tilde b_j({\mathrm y})\partial _{{\mathrm y}_j} + \sum_{j=1}^M  
\tilde B_j e^{{\mathrm y}_{j+1} - {\mathrm y}_j} \nonumber
\end{equation}
$\tilde B_j = B_j - B_{j-1}$. Then the self-adjoint operator ${\cal H}= {1\over 2}(H^{(1)}+ H^{(1)\dag})$ is expressed as
\begin{equation}
{\cal H}= {1\over 2} \sum_{j=1}^M \partial_{{\mathrm y}_j}^2  + \sum_{j=1}^M  \tilde B_j e^{{\mathrm y}_{j+1} - {\mathrm y}_j}, \nonumber
\end{equation}
which is nothing but the Hamiltonian of the Toda chain.

Let us now derive the solution of the set of  SDEs (\ref{SDE1})  introducing suitable integrator factors (we refer the interested reader 
to \cite{Oksendal} on  integrator factors in SDEs). 
Let us consider the general set of SDES
\begin{equation}
d{\mathrm x}_{tj} = b_j({\mathrm x}_t) dt + {\mathrm x}_{tj} d{\mathrm w}_{tj}, \nonumber
\end{equation}
for any drift $b$.
We introduce the following set of integrator factors:
\begin{equation}
{\cal F}_j(t) = \exp\Big ( -\int_0^t d{\mathrm w}_{sj} + {1\over 2}\int_0^t ds \Big ) \label{change} 
\end{equation}
and define the new fields: ${\mathrm y}_{tj} = {\cal F}_{j}(t) {\mathrm x}_{tj}$, then one obtains 
a differential equation for the vector field ${\mathrm y}$. 
Indeed, from (\ref{SDE1}), (\ref{change}) 
\begin{eqnarray}
&&  d( {\cal F}_j(t)^{-1}{\mathrm y}_{tj}) =  d{\mathrm x}_{tj}\  \Rightarrow\    \nonumber\\
&& d{\mathrm y}_{tj} = {\cal F}_{j}(t) b_j\big ({\cal F}^{-1}_k(t){\mathrm y}_{tk}\big) dt \label{solutionf} 
\end{eqnarray}
and we bear  in mind that the usual calculus rules apply for the LHS of the first equation above. 

Let us focus now on the SDEs of the DST model (\ref{SDE1}); in this case (\ref{solutionf})  leads to  the ODE:
\begin{equation}
{d{\mathrm y}_t\over dt}= {\cal A}(t) {\mathrm y}_t, \nonumber
\end{equation}
where the $M \times M$  matrix ${\cal A}$ is given as (we have set $c_j=1$ in (\ref{SDE1}))
\begin{equation}
{\cal A}(t) = {\mathbb I} - {\cal B} = \sum_{j=1}^M \Big ( e_{jj} - {\cal B}_j(t)e_{j j+1}\Big),   \nonumber
\end{equation}
and we define
\begin{equation}
{\cal B}_j(t) = {\cal F}_j(t) {\cal F}^{-1}_{j+1}(t) = \exp \Big ({\mathrm w}_{tj+1} - {\mathrm w}_{tj}\Big ).
\end{equation}
The formal  solution of the latter linear problem is the  path ordered exponential (monodromy):
\begin{equation}
{\mathrm y}_t = {\cal P}\exp\Big ( \int_{0}^t {\cal A}(s) ds \Big ) {\mathrm y}_{0}, \nonumber
\end{equation}
which can be expressed as a formal series expansion:
\begin{eqnarray}
&&{\cal P}\exp\Big ( \int_{0}^t {\cal A}(s) ds \Big )  = \sum_{n=0}^{\infty} \int_0^t \int_0^{t_n} \ldots 
\int_{0}^{t_2} d{t_n} dt_{n-1}  \ldots dt_{1} {\cal A}(t_n) {\cal  A}(t_{n-1}) \ldots {\cal  A}(t_1), \nonumber\\
&&  t\geq t_{n} \geq t_{n-1} \ldots \geq t_{2}.  \nonumber
\end{eqnarray}
Note that ${\cal A}$ is an upper triangular local matrix, thus  products of the matrices preserve the triangular structure, but not the locality. 
In fact, this formal series expansion provides time-like non-local charges of the theory in analogy to the non-local charges associated to representations of 
deformed algebras in quantum and classical integrable models.

It will be instructive to consider the  DST model (\ref{DST0}) and the respective SDEs (\ref{SDE1}) in the continuum limit. 
Let us set $c_j =1$ in (\ref{DST}), then after suitably re-scaling the fields and considering the thermodynamic limit $M \to \infty,\  \delta \to 0$ $(\delta \sim {1\over M}$) we obtain
\begin{eqnarray}
&&  {\mathrm x}_{tj}\  \to\   \varphi(x, t)\nonumber\\
&& {{\mathrm x}_{tj+1} - {\mathrm x}_{tj} \over \delta}\  \to\  \partial_x \varphi(x,t)  \nonumber\\
&& \delta \sum_j  f_j\  \to\  \int dx\ f(x) \nonumber\\
&& {\mathrm w}_{tj}\  \to\  W(x, t). \label{dictionary}
\end{eqnarray}
a brief discussion on the representation of the Brownian sheet $W(x,t)$ (space-time white noise) is presented in the last section.
In the continuum limit the Hamiltonian  of the  DST model (\ref{DST}) becomes the Hamiltonian of an $1+1$ dimensional quantum field theory  
\begin{equation}
H_c^{(1)}= \int dx\ \Big ( {1\over 2} \varphi^2(x) \hat \varphi^2(x) - \partial_x \varphi(x) \hat \varphi(x)\Big ), \nonumber
\end{equation}
where the issue of the ultra locality of underlying algebra has been taken into consideration: $\big [\varphi(x),\ \hat \varphi(y)\big ] = \delta(x-y)$, 
($\hat \varphi(x) = {\partial\over \partial \varphi(x)}$) 
and the SDEs (\ref{SDE1}) become the stochastic transport equation with multiplicative noise:
\begin{equation}
\partial_t \varphi(x,t) = -\partial_x \varphi(x, t) + \varphi(x,t)\dot{W}(x,t) \nonumber
\end{equation}
where $\dot{W}(x,t) = {dW \over dt}$. It is worth noting that in general, the continuum limit of SDEs will lead to non-local SPDEs, 
given that the diffusion matrix is generically a full matrix, and the drift describes in principle non local interactions. However, 
in the case of diagonal (or slightly off diagonal)  diffusion matrices and local drift 
one obtains local SPDEs as the example described above.

\subsubsection*{The DNLS model}

\noindent
We come now to the next member of the DNLS Hierarchy i.e. the quantum DNLS model
(see e.g. \cite{KunduRagnisco} and \cite{ADP} and references therein). 
We directly express the Hamiltonian in terms of differential operators after taking into consideration the map (\ref{map1}):
\begin{equation}
H^{(2)} ={1\over 2} \sum_{j=1}^M \Big (  {\mathrm x}_j ({\mathrm x}_{j+1} -{\mathrm x}_j) \partial_{{\mathrm x}_j}^2 + 
{\mathrm x}_{j+1}^2 \partial_{{\mathrm x}_{j}}\partial_{{\mathrm x}_{j+1}} - 
({\mathrm x}_j -2{\mathrm x}_{j+1} + {\mathrm x}_{j+2})\partial_{{\mathrm x}_j} \Big ) \label{H2}
\end{equation}
and we read the diffusion matrix with entries 
\begin{equation}
g_{jj} ={\mathrm x}_j ({\mathrm x}_{j+1} -{\mathrm x}_j), ~~~~ g_{jj+1} = g_{j+1 j}=  {\mathrm x}_{j+1}^2, ~~~~g_{ij} =0,~~~|i-j |>1,
\end{equation}
as well as drift components $b_j({\mathrm x})=-{1\over 2} ({\mathrm x}_j -2{\mathrm x}_{j+1} + {\mathrm x}_{j+2}) $,
which in the continuum space limit discussed below provides a second derivative of the field.

Using the dictionary described above (\ref{dictionary}) we can readily write down the continuum limit of the DNLS Hamiltonian
\begin{equation}
H_c^{(2)} = {1\over 2} \int dx\  \Big ( \varphi^2(x)\hat \varphi^2(x)- \partial_x^2 \varphi(x) \hat \varphi(x)\Big )
\end{equation} 
as well as the continuum limit of the DNLS SDEs, which is nothing but the stochastic heat equation with multiplicative noise:
\begin{equation}
\partial_t  \varphi(x,t) = -{1\over 2} \partial^2_x \varphi(x, t) + \varphi(x,t)\dot{W}(x,t). \label{heats}
\end{equation}
The latter equation is  solvable,  and can also be mapped to the stochastic viscous Burgers equation  (see for instance \cite{Corwin}). 
Indeed, we set: $\varphi = e^h ,\  u = \partial_x h $ then (\ref{heats}) becomes, 
\begin{eqnarray}
&&  \partial_t  h(x,t) = -{1\over 2} \partial^2_x h(x, t) -{1\over 2}  (\partial_x h(x,t))^2+  \dot{W}(x,t)\nonumber\\
&& \partial_t  u(x,t) = -{1\over 2} \partial^2_x u(x, t) -u(x,t) \partial_x u(x,t)+\partial_x \dot{W}(x,t).
\end{eqnarray}
The latter is precisely the viscous Burgers equation with additive noise.

\subsection{The XXZ quantum spin chain}

\noindent We briefly discuss another prototype integrable model, the XXZ spin ${1\over 2}$ quantum spin chain and the associated SDEs. The XXZ Hamiltonian has the familiar form
\begin{equation}
H = -{1\over 4} \sum_{j=1}^M  \Big ( \sigma_j^x \sigma_{j+1}^x + \sigma_j^y \sigma_{j+1}^y +\Delta \sigma_j^z \sigma_{j+1}^z \Big ) + {\xi \over 8} {\mathbb I}\label{xxx}
\end{equation}
$\sigma^x,\ \sigma^y,\ \sigma^z$ are the familiar $2\times 2$ Pauli matrices, which form a two dimensional representation of the $\mathfrak{sl}_2$ algebra, 
$\Delta =\cosh\mu$, $\mu$ the anisotropy parameter, $\xi$ is an arbitrary constant.  The $\mathfrak{sl}_2$ algebra has generators $S^j$, $j\in \{1,\ 2,\ 3\}$ that satisfy:
\begin{equation}
\big [ S^i,\ S^j \big ] = 2 {\mathrm i} \epsilon_{ijk}S^k. \label{sl}
\end{equation}
The spin $S\in {\mathbb R}$ representation of the algebra in terms of differential operators is given as
\begin{eqnarray}
&& S^1 \mapsto  -({\mathrm x}^2 -1)\partial_{\mathrm x} + S ({\mathrm x}+{\mathrm x}^{-1})\nonumber\\
&& S^2 \mapsto  -{\mathrm i} \Big (({\mathrm x}^2 +1)\partial_{\mathrm x} + S  ({\mathrm x}^{-1}-{\mathrm x})\Big )\nonumber\\
&& S^3 \mapsto 2{\mathrm x}\partial_{\mathrm x}. \label{rep}
\end{eqnarray}
In the case where $S$ is an integer or half integer one obtains the $n=2S+1$ dimensional representation of $\mathfrak{sl}_2$ with a corresponding orthonormal basis 
given as: ${\mathrm e}_{\alpha} = {\mathrm x}^{S-\alpha+1},\ \alpha\in \{1, 2, \ldots, n\}$. This $n$ dimensional space is equipped with an inner product in the unit circle (see also relevant discussion in \cite{DoikouIoannidou}).

Motivated by the form of the XXZ Hamiltonian we replace the Pauli matrices in (\ref{xxx}) with the spin ${1\over 2}$ two dimensional representation of $\mathfrak{sl}_2$ 
expressed in terms of differential operators (\ref{rep}). Then (\ref{xxx}) becomes a typical diffusion reaction Hamiltonian
\begin{eqnarray}
H  &=& \sum_{j} \Big ( {1\over2}\big ({\mathrm x}_{j+1}^2+{\mathrm x} _j^2 -2\Delta {\mathrm x}_j {\mathrm x}_{j+1}\big) \partial_{{\mathrm x}_j} \partial_{{\mathrm x}_{j+1}} +
{1\over 4} \big ({\mathrm x}_j^2 ({\mathrm x}_{j+1}^{-1}+  {\mathrm x}_{j-1}^{-1}) -({\mathrm x}_{j+1} + {\mathrm x}_{j-1}) \big )\partial_{{\mathrm x}_j}  \nonumber\\
&-&{1\over 8}\big({\mathrm x}_j{\mathrm x}_{j+1}^{-1} + {\mathrm x}_j^{-1}{\mathrm x}_{j+1}\big) +{\xi \over 2 }{\mathrm x}^2_j \partial^2_{{\mathrm x}_j}+ 
{\xi \over 2 }{\mathrm x}_j \partial_{{\mathrm x}_j}\Big ) \label{X1} 
\end{eqnarray}
from which one immediately reads the diffusion matrix $g$ as well as the drift $b$, and thus the associated set of SDEs (\ref{st1}).
Recall also $g =\sigma \sigma^T$ one then can identify $\sigma$ and hence $\sigma^{-1}$  in order to apply the results of section 2.  Notice that last two terms in (\ref{X1}) 
correspond to the last term of (\ref{xxx}) after substituting the identity with $\sum_j (\sigma_j^z)^2$, given that $(\sigma_j^z)^2={\mathbb I}_{2\times 2}$.

The  Hamiltonian above can also be expressed in terms of the variables ${\mathrm y}_j,\  \partial_{{\mathrm y}_j}$  (${\mathrm x}_j = e^{{\mathrm y}_j}$) as
\begin{eqnarray}
 H &=&   \sum_j \Big (\cosh{{\mathrm y}_j - {\mathrm y}_{j+1}} -\Delta \Big )\partial_{{\mathrm y}_j}\partial_{{\mathrm y}_{j+1}} + {\xi \over 2}\partial^2_{{\mathrm y}_j}+ 
{\xi \over 2}\partial_{{\mathrm y}_j}\nonumber\\ 
&+& {1\over 2} \sum_j  \Big ( \sinh{{\mathrm y}_j - {\mathrm y}_{j+1}}  + 
\sinh{{\mathrm y}_j - {\mathrm y}_{j-1}} \Big )\partial_{{\mathrm y}_j}
- {1\over 4} \sum_j \cosh{{\mathrm y}_j - {\mathrm y}_{j+1}}.\label{X2} 
\end{eqnarray}
Both expressions (\ref{X1}), (\ref{X2}) may be useful especially when considering the continuum limit of the model
and making connections with certain quantum field theories, this is an important aspect, which however will be studied separately. 
Relevant recent results on stochastic XXZ spin chain are
presented in \cite{Bernard}.

\begin{itemize}

\item{\bf The Ising model}

Two special cases of interest arise as suitable limits 
of the XXZ spin chain:  the XX model when $\Delta =0$ which describes the free fermion point of the sine-Gordon model, and the Ising model when 
$\Delta \to \infty$, allow also $\xi = -2 \hat \xi \Delta$. 
Let us focus  here on the simple case of the Ising model in the presence of an external longitudinal magnetic field, emerging directly from (\ref{X2})
(we have considered the open spin chain for convenience):
\begin{equation}
H = {\hat \xi\over 2} \sum _{j=1}^{M-1}\partial^2_{{\mathrm y}_j}+ {1 \over 2} \sum _{j=1}^{M-1}\partial_{{\mathrm y}_j} 
\partial_{{\mathrm y}_{j+1}} +{\hat \xi\over 2}  \sum_{j=1}^M \partial_{{\mathrm y}_j} + {a \over 2} \partial_{{\mathrm y}_M}^2. \nonumber
\end{equation}
where ${a^2 +1 \over a }= \hat \xi$. The diffusion matrix is now constant and it reads as:
\begin{equation}
g=  \sum_{j=1}^{M-1} \Big (\hat \xi e_{jj} + e_{jj+1} + e_{j+1j} \Big ) +a e_{MM}\  \Rightarrow\  
\sigma = {1\over \sqrt{a}} \Big (a \sum_{j=1}^M e_{jj} + \sum_{j=1}^{M-1} e_{j j+1 }\Big ) \nonumber
\end{equation}
hence the SDEs are given as
\begin{eqnarray}
&& d{\mathrm y}_{tj}= {\hat \xi \over 2}  dt + {1\over  \sqrt{a}}\big  (a\  d{\mathrm w}_{tj} + d{\mathrm w}_{t  j+1}\big ), ~~~~1 \leq j \leq M-1,  \nonumber\\
&& d{\mathrm y}_{tM}= {\hat \xi \over 2}  dt + \sqrt{a}\  d{\mathrm w}_{tM}. \nonumber
\end{eqnarray}
The solutions of the associated SDEs are then easily determined:
\begin{eqnarray}
&& {\mathrm y}_{sj} - {\mathrm y}_{0j} = {\hat \xi \over 2}  s +  {1\over \sqrt{a}}\big ( a\ {\mathrm w}_{sj} +  {\mathrm w}_{sj+1}\big )  ~~~~1 \leq j \leq M-1 \nonumber\\
&& {\mathrm y}_{sM} - {\mathrm y}_{0M} = {\hat \xi \over 2}  s +   \sqrt{a}\ {\mathrm w}_{sM}. \label{Ising}
\end{eqnarray}
The variables ${\mathrm x}_j = e^{{\mathrm y}_j}$ are also immediately identified via the solution above.
Expectation values can readily be computed using the solutions above as described in section 3 (see also the example in section 4.1). 

The addition of a transverse magnetic field, e.g. $h ={ \xi \over 2} \sum_j(S^{1} -{\mathrm i} S^2)$, to the free Hamiltonian  leads to a set of SDEs 
with the same constant diffusion matrix as above, but with drift components proportional to $ e^{{\mathrm y}_{tj}}$. The solution of the associated 
SDEs in this case is a more involved problem,  but it is the first task in order to be able to identify expectation values.

\end{itemize}

As a general remark let us note that in the case of  exactly solvable models, such as for instance the system of $M$-harmonic oscillators, the  DNLS and XXZ models, 
the spectrum and eigenstates of Hamiltonians are available via e.g. the Bethe ansatz formulation 
\cite{Faddeev, Korepin} (see also \cite{Caux}-\cite{Maillet1}). Having a complete basis of eigenstates available one can then express the solution of the time evolution problem 
in terms of this basis. This is rather the standard way for studying the time evolution problem when dealing with quantum mechanical systems with a complete basis of the Hamiltonian eigenstates available. 
However, in the case of non-Hermitian Hamiltonians,  as are the general operators (\ref{general}),  the challenging issue of the existence 
of a complete basis  needs to be addressed. From this point of view the study of the time evolution problem via the path integral approach as described in section 
3  offers a more suitable frame.


\subsection{Quantum Darboux transformation $\&$ defects}

\noindent It will be useful for our purposes here to consider the DST model in the presence of integrable local defects \cite{ADP}, \cite{Corrigan}, \cite{Doikou}. We will derive the associated
quantum Darboux-B\"{a}cklund transform for the DNLS model as a defect matrix and shall also identify the corresponding SDEs. The monodromy in this case is modified as 
(the auxiliary index is suppressed below for brevity)
\begin{equation}
T(\lambda) = {\mathbb L}_{M}(\lambda)\ldots {\mathrm D}_m(\lambda) \ldots {\mathbb L}_{1}(\lambda) \nonumber
\end{equation}
where we consider the defect matrix ${\mathrm D}$ to be of the generic form
\begin{equation}
{\mathrm D}_m(\lambda)=
\begin{pmatrix}
  \lambda +\alpha_m &\beta_m\cr
 \gamma_m & \lambda -\alpha_m
\end{pmatrix}. \nonumber
\end{equation}
The matrix ${\mathrm D}$ will be explicitly derived via the zero curvature condition on the defect point i.e. the $t$ part of the quantum Darboux-B\"{a}cklund transformation \cite{DoikouFindlay}.

In the algebraic scheme preserving integrability requires that ${\mathrm D}$ also satisfies (\ref{algebra}), hence it turns out that $\alpha,\ \beta,\ \gamma$ are the generators of $\mathfrak{sl}_2$. 
Integrability in the weaker sense via the Lax pair description on the other hand leads to the zero curvature condition for ${\mathrm D}$ (the $t$-part of the B\"{a}cklund transformation (\ref{BT}))
\begin{equation}
{d{\mathrm D}_m(\lambda)\over dt}=\tilde  {\mathbb A}_{m}(\lambda){\mathrm D}_m(\lambda)-{\mathrm D}_m(\lambda){\mathbb A}_{m}(\lambda)  \label{BT}
\end{equation}
where $\tilde {\mathbb A}_m$ in general is
\begin{eqnarray}
\tilde {\mathbb A}_m(\lambda) = \begin{pmatrix}
  \lambda  &\tilde z_m\cr
\tilde  Z_{m-1} & 0
\end{pmatrix}, \label{Aop2} \nonumber
\end{eqnarray}
but in our case $\tilde {\mathbb A}_m = {\mathbb A}_{m+1}$.
We first focus on the Lax pair description and we assume that ${\mathbb A}_j$ are given in (\ref{Aop}) for all the sites of the chain, that is we assume the existence of the fields $z_m,\ Z_m$. 
Then solving (\ref{BT}) we obtain the quantum Darboux-B\"{a}cklund relations:
\begin{eqnarray}
&& \beta_m =z_m -\tilde z_{m} \nonumber\\
&& \gamma_m = \tilde Z_{m-1} - Z_{m-1} \nonumber\\
&& \alpha_m^2 = \zeta - \big (z_m-\tilde z_{m}\big )\big (\tilde Z_{m-1}-Z_{m-1}\big ), \label{BT2} 
\end{eqnarray}
where the expression for $\alpha$ follows from requiring the quantum determinant of the matrix ${\mathrm D}$ is a constant. In our case here we treat ${\mathrm D}$ as a defect matrix: 
$\tilde z_m =z_{m+1},\ \tilde Z_{m-1} =Z_m$. 
The corresponding conserved charge in the presence of the local defects then reads as \cite{Doikou}
\begin{eqnarray}
 H &=& {1\over 2} \sum_{j\neq m} z_j^2 Z_j^2 +\Big  (\sum_{j\neq m} c_j z_j - \sum_{j\neq m, m-1}z_{j+1}\Big )Z_j   -z_{m+1}Z_{m-1} \nonumber\\ 
&-&\beta_m Z_{m-1} -\gamma_mz_{m+1}+{\alpha_m^2 \over 2},
\label{ham2b}
\end{eqnarray}
then via relations (\ref{BT2}) and recalling the map (\ref{map1}) the Hamiltonian is rewritten
\begin{equation}
H = {1\over 2} \sum_{j\neq m} {\mathrm x}_j^2 \partial_{ {\mathrm x}_j}^2 +\Big (\sum_{j\neq m}c_j  {\mathrm x}_j - 
\sum_{j\neq m, m-1} {\mathrm x}_{j+1}\Big )\partial {\mathrm x}_j + 
{1\over 2} \Big ( {\mathrm x}_{m+1} \partial_{ {\mathrm x}_{m-1}} - {\mathrm x}_{m} \partial_{{\mathrm x}_{m-1}} -  
{\mathrm x}_{m+1} \partial_{ {\mathrm x}_{m}} -{\mathrm x}_{m} \partial_{{\mathrm x}_m}+\zeta \Big ).  \nonumber
\end{equation}
One can immediately read the corresponding SDEs, indeed for $j\neq m, m-1$ they are given by (\ref{SDE1}) and
\begin{eqnarray}
&& d {\mathrm x}_{tm-1}  = \Big (c_{m-1}  {\mathrm x}_{tm-1}  + {1\over 2}\big ( {\mathrm x}_{tm+1}- {\mathrm x}_{tm}\big )\Big )dt +  
{\mathrm x}_{tm-1} d {\mathrm w}_{tm-1} \nonumber\\
&& d {\mathrm x}_{tm}  =  -{1\over 2}  \Big ({\mathrm x}_{tm+1} +{\mathrm x}_{tm}\Big )dt. \nonumber
\end{eqnarray}
We conclude from the equations above that the $\sigma$ matrix in this case is not invertible as it has one zero eigenvalue, in particular 
$\sigma= \mbox{diag}( {\mathrm x}_1,\  {\mathrm x}_2, \ldots  {\mathrm x}_{m-1},\ 0,\  
{\mathrm x}_{m+1}, \ldots  {\mathrm x}_M)$,
therefore the setting described in subsection 3.2.1  can be implemented.

Let us also employ the algebraic setting, in this case ${\mathrm D}$ is a representation of the algebra (\ref{algebra}), therefore as mentioned 
$\alpha,\ \beta,\ \gamma$ are elements of $\mathfrak{sl}_2$ 
expressed in the Chevalley-Serre basis:
\begin{eqnarray}
\beta ={1\over 2} (S^1 -{\mathrm i}S^2), ~~~~\gamma ={1\over 2} (S^1 + {\mathrm i}S^2), ~~~~\alpha = {1\over 2}S^3. \nonumber
\end{eqnarray}
In the spin $S$ representation (\ref{rep}) they are given by the map:
\begin{eqnarray}
&&\beta_m \mapsto  - {\mathrm x}_n^2\partial_{ {\mathrm x}_m}+S {\mathrm x}_m, \nonumber\\
&&\gamma_m \mapsto \partial_{{\mathrm x}_m}+S {\mathrm x}_m^{-1} \nonumber\\
&&\alpha_m \mapsto  {\mathrm x}_m \partial_{{\mathrm x}_m}, \nonumber
\end{eqnarray}
and the Hamiltonian (\ref{ham2b}) becomes
\begin{eqnarray}
H &=& {1\over 2} \sum_{j\neq m}  {\mathrm x}_j^2 \partial_{x_j}^2 + \Big (\sum_{j\neq m}c_j  {\mathrm x}_j - \sum_{j\neq m, m-1} {\mathrm x}_{j+1}\Big )\partial  {\mathrm x}_j \nonumber\\
&-&  {\mathrm x}_{m+1} \partial_{ {\mathrm x}_{m-1}} +\Big  ( {\mathrm x}_m^2 \partial_{ {\mathrm x}_m} -S {\mathrm x}_m\Big )\partial_{ {\mathrm x}_{m-1}} - \Big (\partial_{ {\mathrm x}_m} + S 
{\mathrm x}_{m}^{-1}\Big ) {\mathrm x}_{m+1} + {1\over 2} 
{\mathrm x}_{m}^2 \partial_{ {\mathrm x}_m}^2 +{1\over 2}  {\mathrm x}_m \partial_{ {\mathrm x}_m}.\nonumber
\end{eqnarray}
The corresponding SDEs are given as before via (\ref{SDE1}) for $j\neq m, m-1$, and
\begin{eqnarray}
&& d {\mathrm x}_{tm-1}  = \Big (c_{m-1}  {\mathrm x}_{tm-1} - {\mathrm x}_{tm+1}-S {\mathrm x}_{tm}\Big )dt +  {\mathrm x}_{tm-1} d {\mathrm w}_{tm-1} +\sqrt{2}  
{\mathrm x}_{tm} d {\mathrm w}_{tm}\nonumber\\
&& d {\mathrm x}_{tm}  =  \Big ({1\over 2}  {\mathrm x}_{tm} - {\mathrm x}_{tm+1}\Big )dt +  {\mathrm x}_{tm}d {\mathrm w}_{tm} + \sqrt{2}  {\mathrm x}_{tm} d {\mathrm w}_{tm-1}. \nonumber
\end{eqnarray}
Notice that the diffusion matrix is not diagonal anymore and interestingly the $\sigma$ matrix is invertible, therefore the quantum canonical transformation can be implemented subject to 
certain modifications associated to the presence of the defect. This model is integrable in the strong sense, and the Bethe ansatz techniques can be applied so the spectrum and 
eigenstates are available. It is also worth noting that in the algebraic setting one obtains deformed ${\mathbb A}$ operators around the defect point:
\begin{equation}
{\mathbb A}_m(\lambda)=
\begin{pmatrix}
  \lambda  &\beta_m+z_{m+1}\cr
 Z_{m-1} &  0
\end{pmatrix},~~~~{\mathbb A}_{m-1} = \begin{pmatrix}
  \lambda  &z_{m+1}\cr
 \gamma_m+Z_{m-1} & 0
\end{pmatrix}. \nonumber
\end{equation}
The operators above  become the usual bulk ones (\ref{Aop}) only in the continuum limit, via analyticity conditions implemented around the defect point. 
These analyticity conditions provide exactly  B\"{a}cklund type relations for the fields \cite{Doikou}. The classical equations of motion, which structurally coincide with the 
quantum ones are available for the model in the presence of  local defects.

\section{Generalizations $\&$ comments}

\noindent 
We may now discuss certain generalizations associated to discrete quantum systems as well as quantum field theories. Let us first consider the obvious extension
from SDEs associated to vector fields ${\mathrm x}$ to SDEs for generic matrix or tensor fields. Indeed,  let us consider the tensor field ${\mathrm Y}$ with components 
${\mathrm Y}_{ i_1 i_2...i_d},\ d \in {\mathbb N}$, then the generator  ${\hat L}_0$  is expressed 
\begin{equation}
{\hat L}_0 =  g_{i_1...i_d; j_1...j_d}({\mathrm Y}){ \partial^2\over \partial {\mathrm Y}_{i_1...i_d}  \partial  {\mathrm Y}_{j_1...j_d}} 
+ b_{i_1...i_d}({\mathrm Y}){ \partial \over \partial {\mathrm Y}_{i_1...i_d}}, ~~~i_k \in \{1, \ldots, M\}. \nonumber
\end{equation}
$g_{i_1...i_d;j_1...j_d}$ and $b_{i_1 ...i_d}$ are the generalized tensor diffusion coefficients and drift components respectively, 
and we use the standard convention where repeated indices are summed.  In the $d=2$ case for instance the generator above is associated to the
partition  function of  a matrix model (see e.g. review articles \cite{TraceyWidom, matrix-models} and references therein, and \cite{AvanJevicki} in relation to integrable models).

The SDEs associated to the generator  -via the generalized It\^{o} formula- are then given as
\begin{equation}
d{\mathrm Y}_{t i_1 ...i_d} = b_{i_1 ...i_d}({\mathrm Y}_t) dt + \sigma_{i_1 ...i_d; j_1...j_d}({\mathrm Y}_t) d{\mathrm w}_{t j_1...j_d} \nonumber
\end{equation}
provided that the tensor Wiener processes satisfy:
\begin{equation}
d{\mathrm w}_{ti_1...i_d}\ d{\mathrm w}_{tj_1...j_d} = \delta_{i_1 j_1} \ldots \delta_{i_d j_d}dt, \nonumber
\end{equation}
and thus $g_{i_1...i_d;j_1...j_d}= \sigma_{i_1...i_d;k_1...k_d}\sigma_{j_1...j_d;k_1...k_d}$. In the continuum limit  ($M \to \infty$), 
which is of particular interest when studying quantum/statistical field theories at finite temperature and SPDEs, 
the tensor fields become continuous space random fields depending on 
$t$ and the continuum space parameters ${\bf x} \in {\mathbb R}^d$, i.e.
\begin{equation}
{\mathrm Y}_{ti_1...i_d}\  \to\  \varphi({\bf x}, t) , ~~~~ {\mathrm w}_{ti_1...i_d}\  \to\  W({\bf x}, t) \nonumber
\end{equation}
and the SDEs become SPDEs; $W({\bf x}, t)$ are the multi-dimensional Wiener fields or Brownian sheets (white space-time noise). 
This description is also in line with the notion  of stochastic quantization in quantum field theory (see for example \cite{stoch-qua1, stoch-qua2}).

We come now to the issue of representing the Brownian sheet (we refer the interested reader to \cite{Prevot}).  
Let $e_k,\  k \in {\mathbb N}$ be an orthonormal basis of a separable Hilbert space ${\cal H}$
consisting of eigen-vectors of an operator $Q$ with corresponding eigenvalues $\lambda_k$. Then the ${\cal H}$ valued stochastic process 
$W_t,\  t \in [0, t]$ (a $Q$-Wiener process)  is represented as
\begin{equation}
W(t) = \sum_{k \in {\mathbb N} } \sqrt{\lambda_k} \beta_k(t) e_k \nonumber
\end{equation}
where $\beta_k(t)$ are independent real valued Brownian motions. Let us present as 
an example the one spatial dimensional case,  then the Wiener field, which is periodic and square integrable in $[-L,\  L]$ is
\begin{equation}
W(x,t) = {\sqrt{L} \over \pi}\sum_{n\geq 1}{1\over n} \Biggl ( X_t^{(n)} \cos{n\pi x \over L}+ Y_t^{(n)} \sin{n\pi x \over L}\Biggr ), \label{sheet1} 
\end{equation}
$X_t^{(n)},\  Y_t^{(n)}$ are independent Brownian motions. In this case we have used as the operator $Q$ the inverse Laplacian: $Q = (\partial_x^2)^{-1}$. 
The representation of higher dimensional Brownian sheets becomes a more complicated issue involving multidimensional Fourier transforms. 
We have already considered two fundamental examples i.e. the DST and DNLS models that in the continuum limit led to the stochastic transport and the 
stochastic heat equations respectively.

The computation of expectation values via the solution of the associated SDEs for the DST (DNLS) and XXZ models is one of our main future goals. 
We have already at our disposal a formal series solution for the DST model, whereas solutions for the XXZ SDEs  (or special cases such as the XX model) 
need to be derived using for instance the change of variables introduced in section 2. 
Moreover, we have discussed the issue of space like defects as a way to tackle the generic case where $\det \sigma =0$, and we have provided a particular example i.e. the DST model
in the presence of local defects. A relevant significant issue is the effect of non-trivial space and time like boundary conditions on the form of the generator of the It\^{o} 
process as well as the form of the SDEs. This case is very interesting especially when one requires  that the boundary conditions preserve the 
integrability of the model.  

Extending the ideas on the construction and solution of non-linear non-local PDEs developed in \cite{BeckDoikouMalhamStylianidis} (i.e. generalizing the notion of the Darboux-dressing
transform) in the case of SPDEs will lead to a modified scheme for producing and solving certain types of  SDEs/SPDEs  \cite{DoikouMalhamWiese}.
Also, the study of the possible connections between the algebraic structures arising  when solving SDEs \cite{FardMalhamWiese}, 
and the deformed algebras associated to integrable systems is  a particularly interesting direction to pursue.

\subsection*{Acknowledgments} 
\noindent A.D. wishes to thank LPTM, University of Cergy-Pontoise, where part of this work was completed, and J. Avan for kind hospitality and useful discussions.


\end{document}